\documentclass[]{pasj01}
\draft

\newcommand{\s}{\mathrm{s}}
\newcommand{\dst}{\mathrm{d}}
\newcommand{\stp}{\mathrm{stop}}

\newcommand{\Msol}{M_{\odot}}
\newcommand{\tstop}{\mathrm{stop}}
\def\vec#1{\mbox{\boldmath $#1$}}

\begin{document} 
\Received{2017/8/7}%{yyyy/mm/dd}
\Accepted{2017/11/16}%{yyyy/mm/dd}
%\Published{yyyy/mm/dd}

\title{Non-linear Development of Secular Gravitational Instability in Protoplanetary Disks}

%%% begin:list of authors
% Do NOT capitalize all letters in "textsc".
\author{Ryosuke T. \textsc{Tominaga}\altaffilmark{1}}%
%\thanks{Example: Present Address is xxxxxxxxxx}}
\altaffiltext{1}{Department of Physics, Nagoya University, Nagoya, Aichi 464-8692, Japan}
\email{tominaga.ryosuke@a.mbox.nagoya-u.ac.jp}

\author{Shu-ichiro \textsc{Inutsuka},\altaffilmark{1}}
%\altaffiltext{2}{B-Address of Institute}
\email{inutsuka@nagoya-u.jp}

\author{Sanemichi Z. \textsc{Takahashi}\altaffilmark{2}\altaffilmark{3}\altaffilmark{4}}
\altaffiltext{2}{Astronomical Institute, Tohoku University, Sendai 980-8578, Japan}
\altaffiltext{3}{Department of Applied Physics, Kogakuin University, Hachioji, Tokyo, 192-0015, Japan}
\altaffiltext{4}{National Astronomical Observatory of Japan, Osawa, Mitaka, Tokyo 181-8588, Japan}
\email{sanemichi@cc.kogakuin.ac.jp}
%%% end:list of authors

%% `\KeyWords{}' always has to be placed before `\maketitle'.
\KeyWords{instabilities --- protoplanetary disks --- methods: numerical } %Do NOT move this preamble from here!

\maketitle

\begin{abstract}
%Please read ``IMPORTANT NOTICE'' carefully before preparing a manuscript. 
We perform non-linear simulation of secular gravitational instability (GI) in protoplanetary disks that has been proposed as a mechanism of the planetesimal formation and the multiple ring formation. Since the timescale of the growth of the secular GI is much longer than the Keplerian rotation period, we develop a new numerical scheme for a long term calculation utilizing the concept of symplectic integrator. With our new scheme, we first investigate the non-linear development of the secular GI in a disk without a pressure gradient in the initial state. We find that the surface density of dust increases by more than a factor of one hundred while that of gas does not increase even by a factor of two, which results in the formation of dust-dominated rings. A line mass of the dust ring tends to be very close to the critical line mass of a self-gravitating isothermal filament. Our results indicate that the non-linear growth of the secular GI provides a powerful mechanism to concentrate the dust. We also find that the dust ring formed via the non-linear growth of the secular GI migrates inward with a low velocity, which is driven by the self-gravity of the ring. We give a semi-analytical expression for the inward migration speed of the dusty ring.  
\end{abstract}

\section{Introduction}\label{sec:intro}

Protoplanetary disks are supposed to be the birth place of planets. Planets are thought to form through the collisional growth from dust grains to planetesimals, which are kilometer-size solid bodies, in a protoplanetary disk. However, the process of the growth from dust to planetesimal is still unclear. Recent high-resolution observations with Atacama Large Millimeter/submillimeter Array (ALMA) have found that a certain class of protoplanetary disks have multiple ring structures (e.g., \cite{HLTau, Isella2016}). The multiple ring structure in HL Tau reported by \citet{HLTau} is a representative example. Various mechanisms are proposed for explaining the ring structures in HL Tau, which include the gravitational interaction between unobserved planets and a disk (e.g., \cite{Kanagawa2015, Akiyama2016}), the dust grain growth including the sintering effects \citep{Okuzumi2016}, and the secular gravitational instability (GI) \citep{Takahashi2014,Takahashi2016}. Since the secular GI has been proposed as not only the formation mechanism of multiple ring-like structures but also that of planetesimals (e.g., \cite{Ward2000,Youdin2005a,Youdin2005b, Youdin2011,Shariff2011,Michikoshi2012,Takahashi2014}), the observed ring structures may be related to the planet formation if they are formed through the secular GI \citep{Takahashi2016}.

The linear growth of the secular GI is well studied. This instability operates even in a self-gravitationally stable disk because of gas-dust friction. For example, \citet{Youdin2011} and \citet{Michikoshi2012} performed the local linear analysis by considering only the equations of dust grains and including the effect of gas turbulence in terms of the velocity dispersion and turbulent diffusion of dust grains. Considering a turbulent disk model, they derived the condition for the growth of the secular GI. \citet{Takahashi2014} first performed the local linear analysis by using equations of both gas and dust. They found that the long-wavelength perturbation, which was thought to be unconditionally unstable in \citet{Youdin2011} and \citet{Michikoshi2012}, is stabilized due to the Coriolis force (see also \cite{Shadmehri2016, Latter2017}). Although these previous works have revealed the physical properties of the linear growth of the secular GI, the non-linear growth has not been studied yet. It is imperative to understand the non-linear growth for describing the planetesimal formation through the secular GI. It is also interesting to determine the surface density contrast in the ring and the gap formed through the non-linear growth of the secular GI since it can be directly compared with actual observations of the multiple ring-like structures. 

In this paper, we study the non-linear growth of the secular GI using a numerical simulation for the first time. The growth timescale of the secular GI is about $10^{2-3}$ times longer than an orbital period of a disk. From the point of view of an accumulation of numerical error, it is very difficult to calculate such a long term evolution with a conventional numerical scheme. For this reason, we develop a new numerical scheme for a long term calculation applying the symplectic method to a scheme for the numerical fluid dynamics. For the first step to understand the non-linear development, we simplify the problem by neglecting the turbulent diffusion, the multiple size distribution, and the growth of dust grains. Moreover, we consider a disk with uniform pressure distribution in order to exclude the radial drift of the dust particles (cf., \cite{Nakagawa1986}) to focus only the property of non-linear growth process. .

This paper is organized as follows. In section \ref{sec:Method} we present our newly developed numerical method for a long term calculation for the secular GI. In section \ref{sec:resultsSGI}, we explain the results of non-linear simulation. 
%Based on the results of our simulation, we discuss the effect of the growth of the secular GI in a disk where the radial drift operates on the planetesimal formation in section \ref{sec:discussion}. 
We find that a dust ring formed via the non-linear growth of the secular GI migrates inward with a low velocity. We give a semi-analytical expression for the inward migration speed of the dusty ring in section \ref{sec:discussion}.
We summarize the conclusion of this work in section \ref{sec:conclusion}.

%\newpage

\section{Method of non-linear calculation}\label{sec:Method}

%In figure \ref{fig:sample}, ...
%
%\begin{figure}
% \begin{center}
%  \includegraphics[width=8cm]{fig1.eps} 
% \end{center}
%\caption{This is the first figure.}\label{fig:sample}
%\end{figure}
%
In this paper, we solve the following hydrodynamic equations for gas and dust with self-gravity and the gas-dust friction under the assumption that a disk is infinitesimally thin and axisymmetric as assumed in the previous works:
\begin{equation}
\frac{\partial \Sigma}{\partial t}+\nabla\cdot\left(\Sigma \vec{u} \right)=0, \label{eocgas}
\end{equation}

\begin{eqnarray}
\Sigma\left(\frac{\partial \vec{u}}{\partial t}+\left(\vec{u}\cdot\nabla \right)\vec{u} \right)&=&-\nabla P-\Sigma\nabla\left(\Phi-\frac{G M_{\ast}}{r} \right)\nonumber\\
&&\;\;\;+\frac{\Sigma_{\mathrm{d}}\left(\vec{v}-\vec{u} \right)}{t_{\mathrm{stop}}},\label{eomgas}
\end{eqnarray}

\begin{equation}
\frac{\partial \Sigma_{\mathrm{d}}}{\partial t}+\nabla\cdot\left(\Sigma_{\mathrm{d}} \vec{v} \right)=0,\label{eocdust}
\end{equation}

\begin{eqnarray}
\Sigma_{\dst}\left(\frac{\partial \vec{v}}{\partial t}+\left(\vec{v}\cdot\nabla \right)\vec{v} \right)&=&-c_{\dst}^2\nabla\Sigma_{\dst}-\Sigma_{\dst}\nabla\left(\Phi-\frac{G M_{\ast}}{r} \right)\nonumber\\
&&\;\;\;+\frac{\Sigma_{\mathrm{d}}\left(\vec{u}-\vec{v} \right)}{t_{\mathrm{stop}}},\label{eomdust}
\end{eqnarray}

\begin{equation}
\nabla^2\Phi=4\pi\mathrm{G}\left(\Sigma+\Sigma_{\dst} \right)\delta(z),\label{poisson}
\end{equation}
where $\Sigma$ and $\Sigma_{\dst}$ represent the surface densities of gas and dust, $\vec{u}$ and $\vec{v}$ denote the velocities of those components. The gas pressure is denoted by $P$. We use $M_{\ast}$ to represent the mass of the central star. The stopping time of a dust particle is denoted by $t_{\stp}$. Equations (\ref{eocgas}) and (\ref{eomgas}) are the equation of continuity and the equation of motion for gas. Equations (\ref{eocdust}) and (\ref{eomdust}) are those for dust. In this work, we consider the effect of the velocity dispersion of dust particles $c_{\dst}$ using the first term in the right hand side of equation (\ref{eomdust}). Equation (\ref{poisson}) is the Poisson equation for the self-gravity of gas and dust disks.
\begin{figure}[ht!]
%\centering
\begin{center}
\hspace{0pt}
\includegraphics[width=8cm]{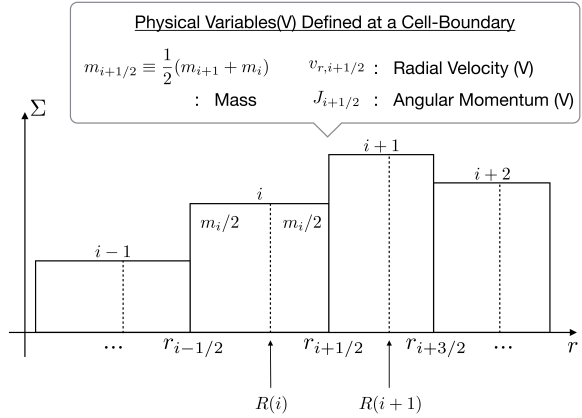}
\end{center}
\caption{Positions where variables are defined. The surface density is denoted by $\Sigma$. We use $r_{i+1/2}$ to denote a radius of the boundary between the $i$-th and the $(i+1)$-th cells, on which the radial velocity and the angular momentum are defined, and $R(i)$ to denote a radius that divides the mass equally in the $i$-th cell. \label{fig:f1m}}
\end{figure}

Since the growth timescale of the secular GI is orders-of-magnitudes longer than the Keplerian period, even a small numerical error created in a Keplerian timescale tends to be excessively accumulated in time integration over a growth timescale of the secular GI.  Thus, to obtain a meaningful numerical result we have to use extremely accurate scheme for time integration. For this purpose, we develop a special numerical scheme for accurate time integration for computational hydrodynamics. We first develop a new one-dimensional code based on the concept of the Lagrangian description of hydrodynamics by applying the symplectic integration method for a Hamiltonian system. In this scheme, we divide a gaseous disk into $N$ fluid elements, and update the time evolution of the positions of those fluid elements. Because the equation of motion of the fluid element in this scheme does not include the advection term, the method is free from dissipation due to numerical advection. Figure \ref{fig:f1m} shows positions where physical variables are defined in cylindrical coordinates. Our centering of variables is apparently analogous to that of staggered mesh method (e.g., \cite{Stone1992}), but conceptually different in the following sense: in figure 1 the mass is conserved in the cell bounded by solid lines, and the velocity is regarded as the average velocity for the two adjacent rectangles bounded by the dashed lines. We use the same structure of cells for dust component of the disk. In the absence of the friction between gas and dust, our new scheme is a symplectic scheme that is free from numerical dissipation and can be regarded as a discretized Hamiltonian system, which enables very accurate long term numerical integration (see section \ref{sec:gd-friction}). In appendix \ref{sec:ap1}, the formulation of our scheme is explained in detail. In this work, we use the leap-frog method, which is second-order time integrator.

We include the self-gravity and the gas-dust friction in our symplectic method, and do numerical calculations of the secular GI. Although the friction term makes the system non-Hamiltonian, we can accurately describe the effect of dissipation caused by the friction term because our basic scheme is free from other numerical dissipation.  We describe the self-gravity solver, result of test calculation of standard GI, and a method to calculate the gas-dust friction in the following sections.

%--------2.1------------------------
\subsection{Self-gravity solver}\label{subsec:sg}

In this study, we calculate the self-gravity by summing up the gravitational forces from infinitesimally thin rings, which satisfy the following conditions: in cylindrical coordinate $(r,\theta,z)$, (i) the center of rings is located at the origin of coordinate, (ii) all rings are in the z=0 plane, and (iii) these have a uniform line density. When a ring mass is  $M_{\mathrm{ring}}$ and radius is $a$, gravitational potential $\Phi_{\mathrm{ring}}(r,z;a)$ at a point $(r,\theta,z)$ is given by
\begin{equation}
\Phi_{\mathrm{ring}}(r,z;a) = M_{\mathrm{ring}}U(r,z;a),
\end{equation}
\begin{equation}
U(r,z;a)\equiv-\frac{2G K(n)}{\pi p},\label{ringpot}
\end{equation}
where $G$ is gravitational constant, and $p$ and $n$ are defined as follows:
\begin{equation}
p\equiv\sqrt{(r+a)^2+z^2},  
\end{equation}
\begin{equation}
n\equiv\frac{4ar}{p^2}. 
\end{equation}
The function $K(n)$ is the complete elliptic integral of the first kind
\begin{equation}
K(n)\equiv\int^{\pi/2}_0 \frac{d\theta}{\sqrt{1-n\sin^2\theta}} .
\end{equation}
The ring gravity per unit mass $F_r(r,z;a)$ is obtained by differentiating the potential and written as
\begin{equation}
F_r(r,z;a)\equiv M_{\mathrm{ring}}\tilde{F}(r,z;a)
\end{equation}
\begin{equation}
\tilde{F}(r,z;a)=-\frac{G}{\pi p}\left[\frac{K(n)}{r}+A(r,z;a)\frac{E(n)}{q^2} \right],\label{ringgrav}
\end{equation}
\begin{equation}
A(r,z;a)\equiv\frac{r^2-z^2-a^2}{r},
\end{equation}
\begin{equation}
q\equiv\sqrt{(r-a)^2+z^2},
\end{equation}
where $E(n)$ is the complete elliptic integral of the second kind
%----------------------------------
\begin{equation}
E(n)\equiv\int^{\pi/2}_0\sqrt{1-n\sin^2\theta}d\theta.
\end{equation}

We approximate the self-gravity per unit mass at $r = r_{i+1/2}$ with using the following $\delta m_{i+1/2}$ defined in terms of surface density $\Sigma$ and unperturbed surface density $\Sigma_0$ as follows:
\begin{equation}
\delta m_{i+1/2}\equiv \frac{\delta m_{i+1} + \delta m_i}{2},
\end{equation}
\begin{equation}
\delta m_{i}\equiv \pi\left(\Sigma - \Sigma_0 \right)\left(r^2_{i+1/2}-r_{i-1/2}^2\right),
\end{equation}
\begin{equation}
\sum_{i\ne j}\delta m_{j+1/2}\tilde{F_r}(r_{i+1/2},0;r_{j+1/2}).\label{sggrav}
\end{equation}
Above method with $\delta m$ corresponds to the solution of the Poisson equation with perturbed surface density $\delta \Sigma\equiv\Sigma-\Sigma_0$ as a source term. 
In this work, we use approximate functions to evaluate the elliptic integrals in equation (\ref{sggrav}) \citep{Hastings1955}. When we use softening length $h$ for the self-gravity, we use the following equation
\begin{equation}
\sum_{i\ne j}\delta m_{j+1/2}\tilde{F}_r(r_{i+1/2},h;r_{j+1/2}).
\end{equation}
We also use a correction term for the self-gravity to increase the accuracy, which corresponds to the self-gravity from its own cell (see, appendix \ref{sec:ap2}). 

%----------------2.2-------------------
\subsection{Hydrodynamics solver for non-dissipative system}\label{subsec:hsolv}
We apply the symplectic method to the numerical calculation of fluid dynamics. The symplectic method is time integrator especially used for the calculation of a long term evolution of a Hamiltonian system. The main feature is the absence of secular increase of errors in energy. We combine this integrator and the finite volume method. We calculate force exerting on cell-boundaries, and solve time evolution of these position. Our Lagrangian method without advection term in the equation of motion is in remarkable contrast to usual Eulerian methods that unavoidably introduce dissipation due to smoothing of variables caused by advection. The formulation is described in detail in appendix \ref{sec:ap1}.

With the self-gravity solver described in section \ref{subsec:sg}, we test our scheme by calculating time evolution of self-gravitationally unstable disk. We consider a disk rotating around a solar mass star with the Keplerian velocity. The center of the domain is set at $r=80$ au, and the width of the domain $L$ is set to be twice the most unstable wavelength. We use a piecewise polytropic relation (cf., \cite{Machida2006})
\begin{equation}
P=c_{\s,0}^2\Sigma_0\left[ \frac{\Sigma}{\Sigma_0} + \left( \frac{\Sigma}{\Sigma_0}\right)^{5/3}\right],\label{eos}
\end{equation}
where $c_{\s,0}$ is isothermal sound speed and we assume $c_{\s,0}\simeq0.18$ m $\mathrm{s}^{-1}$. This sound speed is a thousand times smaller than a sound speed with mean molecular weight 2.37 and temperature 10K. By assuming this small sound speed, the ratio of the most unstable wavelength to disk radius becomes of the order of $10^{-4}$, and we can directly compare numerical results with the results of the local linear analysis. 
\begin{figure}[ht]
\centering
%\hspace{-70pt}
%\begin{center}
%\includegraphics[bb=0 0 288 216,clip,width=\hsize]{./f2.pdf}
%\includegraphics[bb=0 0 288 216,clip,width=8cm]{./f2.pdf}
%\FigureFile(8cm,8cm){f2.eps}
\hspace{-65pt}\raisebox{-220pt}[0pt][200pt]{\includegraphics[width=8cm]{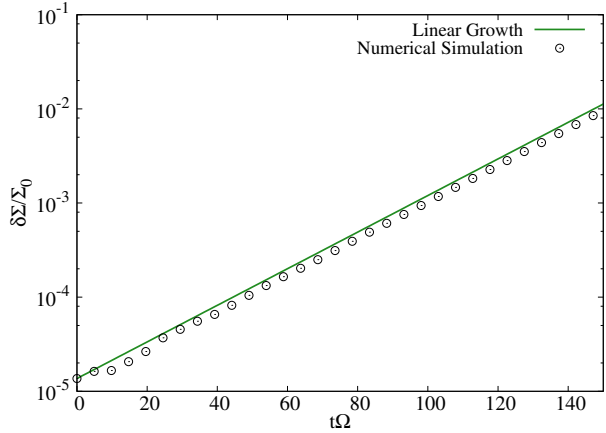}}
%\includegraphics[clip,width=8cm]{./f2.eps}
%\end{center}
%\vspace{-10pt}
\caption{The result of calculation of self-gravitationally unstable disk. The horizontal axis is time normalized by angular velocity, and the vertical axis is the amplitude of the surface density perturbation normalized by the initial surface density at the center of the domain. The black circles represent the result of numerical calculation, the solid green line represents the linear growth rate. In this calculation, the unperturbed density is set for Toomre's $Q$ to be 0.999, and initial perturbation is given by eigenfunction of the most unstable mode obtained by using local linear analysis. The growth rate is about $4.5\times10^{-2}\Omega$. This figure shows that, with our new scheme, we can accurately calculate such a long time evolution. \label{fig:f1}}
\end{figure}
We set the unperturbed surface density $\Sigma_0$ for Toomre's $Q$ value (equation \ref{toomre})\citep{Toomre1964} to be 0.999 at $r=80$ au,
\begin{equation}
Q=\frac{c_{\s}\Omega}{\pi G\Sigma_0}\label{toomre},
\end{equation}
where $\Omega$ is angular velocity of the disk, $c_{\s}\equiv\sqrt{8/3}c_{\s,0}$ is sound speed. Initial perturbation is given by the eigenfunction of the most unstable mode obtained by using the local linear analysis with the amplitude $\delta \Sigma/\Sigma_0\sim10^{-5}$. We use the fixed boundary condition. We give a density distribution outside the domain, which is given by the eigenfunction of the surface density. The amplitude of the outside surface density distribution is increased with the growth rate at the center of the domain. Both sides of the external density fields are resolved with 128 cells respectively. The number of cells in the domain is 512. Time interval $\Delta t $ is set to be $\Delta t =L/512c_{\s}$. We use softening length $h$ and fix this value $h=L/4N$. Figure \ref{fig:f1} shows the time evolution of the density at the center of the domain. The maximum growth rate is about $4.5\times 10^{-2}\Omega$ when Toomre's $Q$ values is equal to 0.999, which is much slower than the Keplerian rotation rate. We can accurately calculate this very slow evolution with our symplectic method.
%\begin{figure}[ht!]
%\centering
%\includegraphics[bb=0 0 288 216,clip,width=\hsize]{./f2.pdf}
%\caption{The result of calculation of self-gravitationally unstable disk. The horizontal axis is time normalized by angular velocity, and the vertical axis is the amplitude of the surface density perturbation normalized by the initial surface density at the center of the domain. The black circles represent the result of numerical calculation, the solid green line represents the linear growth rate. In this calculation, the unperturbed density is set for Toomre's $Q$ to be 0.999, and initial perturbation is given by eigenfunction of the most unstable mode obtained by using local linear analysis. The growth rate is about $4.5\times10^{-2}\Omega$. This figure shows that, with our new scheme, we can accurately calculate such a long time evolution. \label{fig:f1}}
%\end{figure}

%---------------2.3---------------------------
\subsection{Gas-dust friction}\label{sec:gd-friction}
In this section, we show how to solve the gas-dust friction terms in the equation of motion. There are two timescales that is important to describe the motion of dust and gas in a protoplanetary disk: one is an orbital period of the disk $\simeq \Omega^{-1}$, the other is the stopping time $t_{\stp}$. The product $t_{\stp}\Omega$ becomes much smaller than unity when we consider the motion of small dust particles. In this case, we must set $\Delta t$ much smaller than $t_{\stp}$ to conduct a stable calculation, while the growth timescale is much longer than the orbital period. Therefore, the use of such a small time step should be a disadvantage in a numerical calculation of the secular GI.

In this work, we use the piecewise exact solution for the gas-dust friction term \citep{Inoue2008}. The piecewise exact solution is an operator splitting method and used to integrate numerically a stiff differential equation. Since this method is unconditionally stable, we can conduct a calculation without the restriction of $\Delta t$ due to small $t_{\stp}$. With our symplectic scheme, based on the concept of Lagrangian description,  we update a time evolution of each fluid element. In this case, a position of dust cell does not necessarily coincide with that of gas cell. We show the way to apply the piecewise exact solution to our symplectic method in the following. First, we interpolate physical values in cells with certain functions. The interpolation functions are shown in appendix \ref{sec:ap3} in detail. Next, using the interpolation functions in an overlapped region of the $j$-th dust cell and the $i$-th gas cell, we integrate each physical value and calculate masses of each fluid $m^k_{\mathrm{g}},m^k_{\dst}$, radial linear momentums $P^k_{\mathrm{g}},P^k_{\dst}$, and angular momentums $J^k_{\mathrm{g}},J^k_{\dst}$ (see figure \ref{fig:pexact}).  We update time evolution of each momentum with momentum changes due to the friction term:
\begin{figure}[ht!]
%\centering
\begin{center}
%\includegraphics[bb=0 0 512 389,clip,width=\hsize]{./f3.pdf}
%\includegraphics[bb=0 0 512 389,clip,width=8cm]{./f3.pdf}
%\FigureFile(8cm,8cm){f2.eps}
\includegraphics[clip,width=8cm]{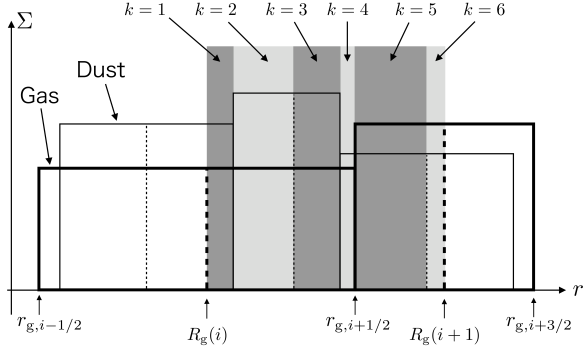}
\end{center}
\caption{Regions used for the piecewise exact solutions (colored with dark and light gray). The boundary of the regions is defined by the boundary or the center of gas and dust cells. The cells enclosed by thick lines represent gas cells, and the cells enclosed by thin lines represent dust cells. In this figure, we consider regions, which are colored with dark and light gray, related to momentum changes at the boundary between the $i$-th and the $(i+1)$-th cells of gas component (equations (\ref{anapg}) and (\ref{anajg})). The number $k$ characterizes each region. \label{fig:pexact}}
\end{figure}
\begin{equation}
\frac{d P_{\mathrm{g}}^k}{d t}=-m^k_{\dst}\frac{U^k-V^k}{t^k_{\mathrm{stop}}},
\end{equation}
\begin{equation}
\frac{d P^k_{\dst}}{d t}=-m^k_{\dst}\frac{V^k-U^k}{t^k_{\mathrm{stop}}},
\end{equation}
\begin{equation}
\frac{d J^k_{\mathrm{g}}}{d t}=-m^k_{\dst}\frac{l^k_{\mathrm{g}}-l^k_{\dst}}{t^k_{\mathrm{stop}}},
\end{equation}
\begin{equation}
\frac{d J^k_{\dst}}{d t}=-m^k_{\dst}\frac{l^k_{\dst}-l^k_{\mathrm{g}}}{t^k_{\mathrm{stop}}},
\end{equation}
where $U^k\equiv P^k_{\mathrm{g}}/m^k_{\mathrm{g}}$ and $V^k\equiv P^k_{\dst}/m^k_{\dst}$ denote radial velocity, and $l^k_{\mathrm{g}}\equiv J^k_{\mathrm{g}}/m^k_{\mathrm{g}}$ and $l^k_{\dst}\equiv J^k_{\dst}/m^k_{\dst}$ represent specific angular momentum. Assuming that $m^k_{\mathrm{g}}$ and $m^k_{\dst}$ are constant, we obtain the analytical solution of these differential equations as follows:
\begin{equation}
P^k_{\mathrm{g}}(t+\Delta t)=P^k_{\mathrm{g}}(t)-\left[U^k(t)-V^k(t) \right]f^k(\Delta t),
\end{equation}
\begin{equation}
P^k_{\dst}(t+\Delta t)=P^k_{\dst}(t)+\left[U^k(t)-V^k(t) \right]f^k(\Delta t),
\end{equation}
\begin{equation}
J^k_{\mathrm{g}}(t+\Delta t)=J^k_{\mathrm{g}}(t)-\left[l^k_{\mathrm{g}}(t)-l^k_{\dst}(t) \right]f^k(\Delta t),
\end{equation}
\begin{equation}
J^k_{\dst}(t+\Delta t)=J^k_{\dst}(t)+\left[l^k_{\mathrm{g}}(t)-l^k_{\dst}(t) \right]f^k(\Delta t),
\end{equation}
\begin{equation}
f^k(t)\equiv \frac{m^k_{\mathrm{g}}m^k_{\dst}}{m^k_{\dst}+m^k_{\mathrm{g}}}\left[1-\exp\left( \frac{m^k_{\dst}+m^k_{\mathrm{g}}}{m^k_{\mathrm{g}}}\frac{t}{t^k_{\mathrm{stop}}}\right) \right],
\end{equation}
where we integrate from $t$ to $t+\Delta t$. We update the momentums and the angular momentums with the analytical solution for each region where a dust cell overlaps with a gas cell. Finally, we sum up updated values $P^k_{\mathrm{g}}$, $P^k_{\dst}$, $J^k_{\mathrm{g}}$, and $J^k_{\dst}$ in each cell, and we calculate radial linear momentums of a cell boundary, $P^k_{\mathrm{g},i+1/2}$ and $P^k_{\dst,i+1/2}$, and angular momentums, $J^k_{\mathrm{g,i+1/2}}$ and $J^k_{\dst,i+1/2}$:
\begin{equation}
P_{\mathrm{g},i+1/2}(t+\Delta t)=\sum_{k}P^k_{\mathrm{g}}(t+\Delta t),\label{anapg}
\end{equation}
\begin{equation}
P_{\dst,i+1/2}(t+\Delta t)=\sum_{k}P^k_{\dst}(t+\Delta t),\label{anapd}
\end{equation}
\begin{equation}
J_{\mathrm{g},i+1/2}(t+\Delta t)=\sum_{k}J^k_{\mathrm{g}}(t+\Delta t),\label{anajg}
\end{equation}
\begin{equation}
J_{\dst,i+1/2}(t+\Delta t)=\sum_{k}J^k_{\dst}(t+\Delta t),\label{anajd}
\end{equation}
With using this method, total linear momentum and total angular momentum conserve exactly.

The actual procedure of numerical calculations for the secular GI is as follows:
\begin{enumerate}
\item Update positions and radial velocities of gas cells and dust cells with forces without the friction term with the symplectic integrator,
\item Update radial linear momentums and angular momentums using equations (\ref{anapg})-(\ref{anajd}),
\item Update radial velocities of gas cells and dust cells with updated momentums and the following relations, 
\begin{eqnarray}
u_{r,i+1/2}=P_{\mathrm{g},i+1/2}/m_{\mathrm{g},i+1/2}\\
v_{r,i+1/2}=P_{\dst,i+1/2}/m_{\dst,i+1/2}
\end{eqnarray}
where $i=1,2,3,...,N-1$. We use the radial velocities and the angular momentums obtained in step 2 and 3 in step 1 in the next time integration.
\end{enumerate}
%\begin{figure}[ht!]
%\centering
%\includegraphics[bb=0 0 512 389,clip,width=\hsize]{./f3.pdf}
%\caption{Regions used for the piecewise exact solutions (colored with yellow and orange). The boundary of the regions is defined by the boundary or the center of gas and dust cells. The red cells represent gas cells, and the blue cells represent dust cells. In this figure, we consider regions, which are colored with yellow and orange, related to momentum changes at the boundary between the $i$-th and the $(i+1)$-th cells of gas component (equations (\ref{anapg}) and (\ref{anajg})). The number $k$ characterizes each region. \label{fig:pexact}}
%\end{figure}

\section{Results of non-linear evolution}\label{sec:resultsSGI}

%In the theory (\cite{key-1})..........
%
%\subsection{Subsection}
%
%In the theory (\cite{key-2})..........
%
%\subsubsection{Subsubsection}
%
%The resent result from ...........
%------------------3.1---------------------------
\subsection{Comparison with the local linear analysis}\label{sec:complinearana}
We test our scheme described in section \ref{sec:Method} by calculating the time evolution of the linear growth of the secular GI. We consider a disk rotating around a solar mass star with the Keplerian velocity. The center of the domain is set at $r=100$ au, and the width of the domain $L$ is set to be the quadruple of the most unstable wavelength. We use the piecewise polytropic relation for gas (equation (\ref{eos})). We assume $c_{\s,0}\simeq$ 1.8 m $\mathrm{s}^{-1}$. We use the isothermal equation of state for dust $P=c^2_{\dst}\Sigma_{\dst}$ assuming $c_{\dst}=0.1c_{\s}=0.1\sqrt{8/3}c_{\s,0}$. By assuming this small sound speed, the ratio of the most unstable wavelength to the disk radius becomes of the order of $10^{-3}$ so that we can directly compare numerical results with the results of the local linear analysis. We set the unperturbed surface density of dust and gas for Toomre's $Q$ value of each fluid to be 3 at $r=100$ au. Initial perturbation is given by the eigenfunction of the most unstable mode obtained by using the local linear analysis with the amplitude $\delta \Sigma/\Sigma_0\sim10^{-4}$. We use the fixed boundary condition. To calculate the gravitational force, we give the surface density distribution outside the domain, which is given by the eigenfunction of the surface density. The amplitude of the outside surface density distribution is increased with the growth rate at the center of the domain. Both sides of the external density fields are resolved with 64 cells. The number of cells in the domain is 512. The time interval $\Delta t$ is given by the following equation at each cell, and we use the minimum value:
\begin{equation}
\Delta t_{i} = 0.5\left(\frac{r_{i+1/2}-r_{i-1/2}}{C_{i}}\right),
\end{equation}
\begin{equation}
C_{i}\equiv c_i+\sqrt{\left|\left(r_{i+1/2}-r_{i-1/2} \right)\frac{\partial \Phi}{\partial r}|_{r=r_{i+1/2}}\right|},
\end{equation}
where $c_i$ represent $c_{\mathrm{s}}$ or $c_{\dst}$ at the $i$-th cell. We use a variable softening length. Initially we set the softening length $h_{i+1/2}=L/4N$ at $r=r_{i+1/2}$. In every time-step, we adopt the following value for the region where the surface density perturbation is positive:
\begin{eqnarray}
h_{i+1/2}&=&\frac{1}{4}(\Delta r_{i+1}+\Delta r_{i}),\\
&=&\frac{1}{8}(r_{i+3/2}-r_{i-1/2}),
\end{eqnarray}
\begin{figure}[ht!]
%\centering
\begin{center}
%\includegraphics[bb = 0 0 288 216, width=\hsize]{./f4.pdf}
%\includegraphics[bb = 0 0 288 216, width=8cm]{./f4.pdf}
%\FigureFile(8cm,8cm){f2.eps}
\hspace{-65pt}\raisebox{-225pt}[0pt][200pt]{\includegraphics[width=8cm]{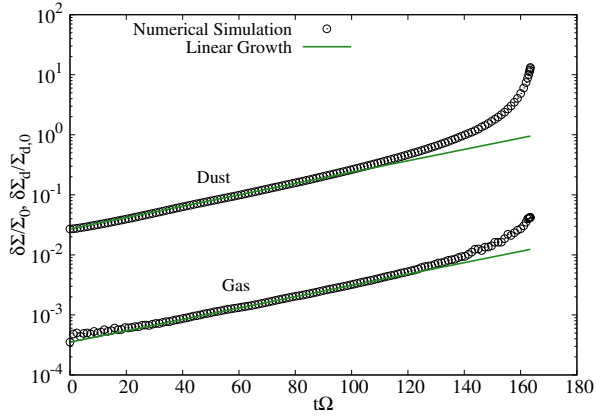}}
\end{center}
\caption{Results of the test calculation of the secular GI in the case of $t_{\tstop}\Omega=1$. The horizontal axis is time normalized by the angular velocity, and the vertical axis is the amplitude of the surface density perturbation normalized by the initial surface density at the center of the domain. The black circles represent the result of numerical calculation, and the solid green lines represent the linear growth. In this case, the growth rate is of the order of $10^{-2}$. Even in the presence of the gas-dust friction, we can accurately calculate the evolution with our symplectic scheme. \label{fig:ap4f1}}
\end{figure}
\begin{figure}[htbp]
%\hspace{-70pt}\raisebox{0pt}[0pt][0pt]{
	\begin{tabular}{c}
		\begin{minipage}{0.48\hsize}
			\begin{center}
				\hspace{-65pt}\raisebox{-225pt}[0pt][200pt]{\includegraphics[width=8cm]{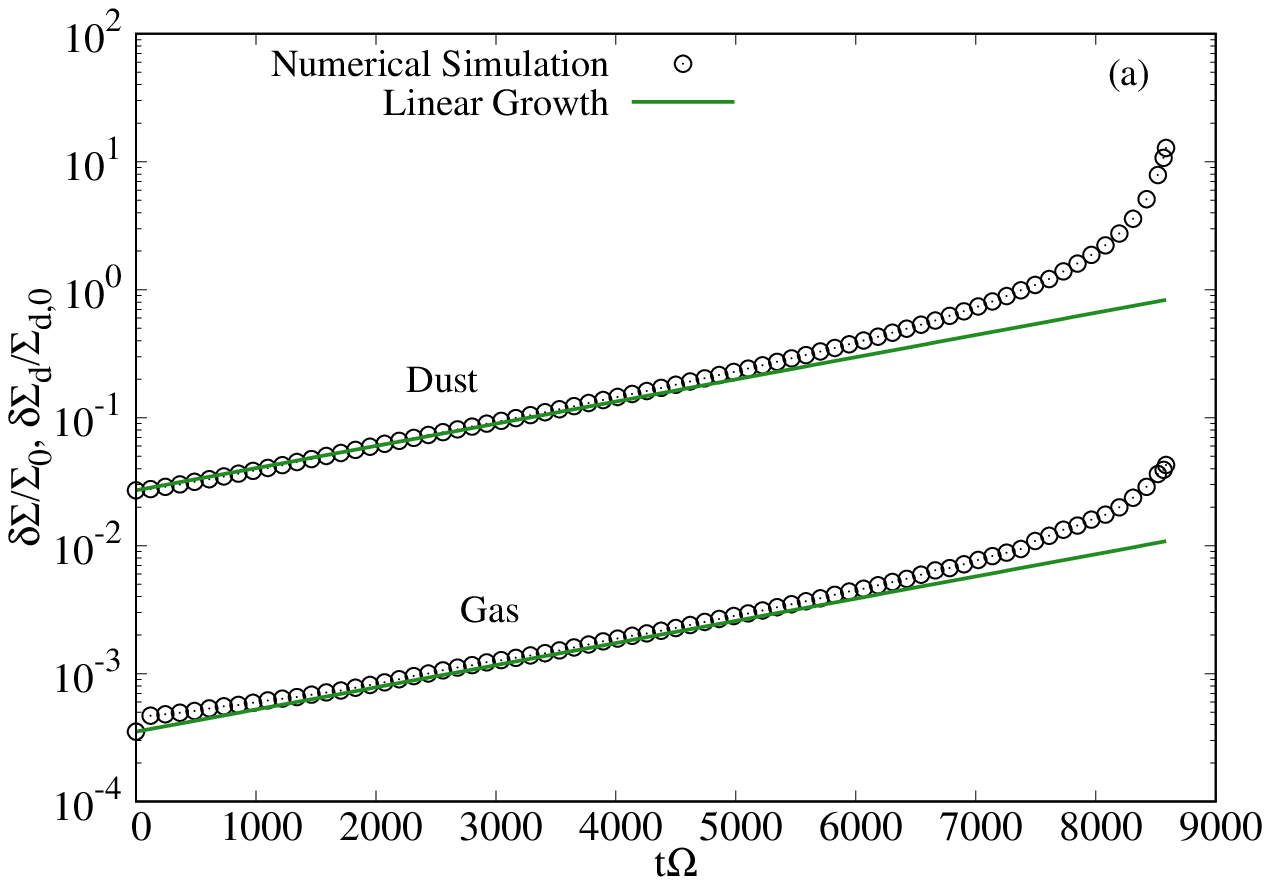}}
			\end{center}
		\end{minipage}
		\begin{minipage}{0.48\hsize}
			\begin{center}
				\hspace{-65pt}\raisebox{-225pt}[0pt][200pt]{\includegraphics[width=8cm]{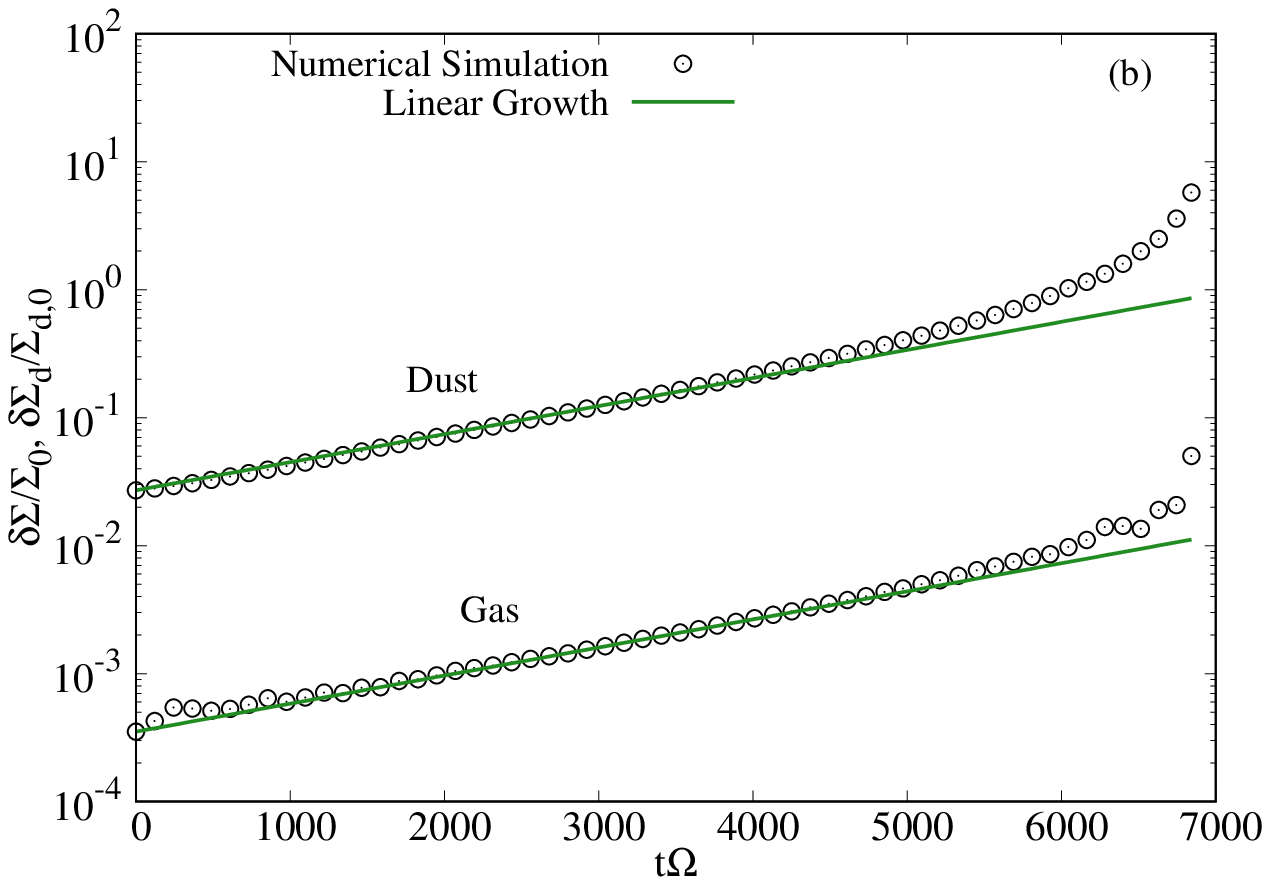}}
			\end{center}
		\end{minipage}
	\end{tabular}
%	}
\vspace{15pt}
\caption{Results of the test calculation of the secular GI in the cases of $t_{\tstop}\Omega=0.01, 100$. The horizontal axis is time normalized by the angular velocity, and the vertical axis is the amplitude of the surface density perturbation normalized by the initial surface density at the center of the domain. The black circles represent the result of numerical calculation, and the solid green lines represent the linear growth. The left figure shows the result for $t_{\tstop}\Omega=0.01$, and the right figure shows that for $t_{\tstop}\Omega=100$. In these cases, the growth rates are of the order of $10^{-4}$. We can accurately calculate these very slow evolution with our symplectic scheme.\label{fig:ap4f2}}
\end{figure}
where $\Delta r_i$ denotes the width of the $i$-th cell. Reducing the softening length for the high surface density region, we evaluate the self-gravity in the infinitesimally thin disk correctly. We consider three cases, $t_{\tstop}\Omega=0.01, 1$, and 100. Figures \ref{fig:ap4f1}  and  \ref{fig:ap4f2} show the time evolution of the surface density at the center of the domain. 
The maximum growth rates in the cases of $t_{\tstop}\Omega=0.01$ and 100 are of the order of $10^{-4}$. Even in the presence of the gas-dust friction, we can accurately calculate these very slow evolution with our symplectic scheme. In every case, the growth deviates from the linear growth of the secular GI from the time when $\delta \Sigma_{\dst}/\Sigma_{\dst,0}$ reaches about 0.3. The dust surface density tends to grow infinitely in the limit of vanishing softening length, which means that the dust ring collapses within a finite time. We estimate the time of the collapse of the dust ring with the following way. First, fitting the surface density of dust that grows non-linearly using a function $f(\tilde{t})\propto(\tilde{t}_{\mathrm{c}}-\tilde{t})^{-q}$, where $\tilde{t}\equiv tn_{\mathrm{SGI}}$ represents time normalized by the linear growth rate of the secular GI $n_{\mathrm{SGI}}$, we estimate the collapse time $\tilde{t}_{\mathrm{c}}$ (figure \ref{fig:fittaus1e0}). On the other hand, we define “the linear theory"-based surface density at $t=t_{\mathrm{c}}$ as a result of hypothetical exponential growth, $\delta \Sigma_{\dst} (t=t_{\mathrm{c}}) = \delta \Sigma_{\dst} (t=0)\exp(n_{\mathrm{SGI}}t_{\mathrm{c}})$. Using this, we define $\delta_{\mathrm{c}}=\delta \Sigma_{\dst}(t=t_{\mathrm{c}})/\Sigma_{\dst,0}$ as the instant of gravitational collapse. In order to study the dependence of $\delta_{\mathrm{c}}$ on the strength of the dust-gas friction, we estimate $\delta_{\mathrm{c}}$ with varying the stopping time. We obtain $\delta_{\mathrm{c}}\simeq 1$ for cases with $t_{\tstop}\Omega=0.01, 0.1, 1, 10,$ and 100 (figure \ref{fig:deltac-taus}). 
\begin{figure}[ht!]
%\centering
\begin{center}
%\includegraphics[bb = 0 0 288 216, width=\hsize]{./f6.pdf}
%\includegraphics[bb = 0 0 288 216, width=8cm]{./f6.pdf}
%\FigureFile(8cm,8cm){f2.eps}
\hspace{-65pt}\raisebox{-225pt}[0pt][200pt]{\includegraphics[width=8cm]{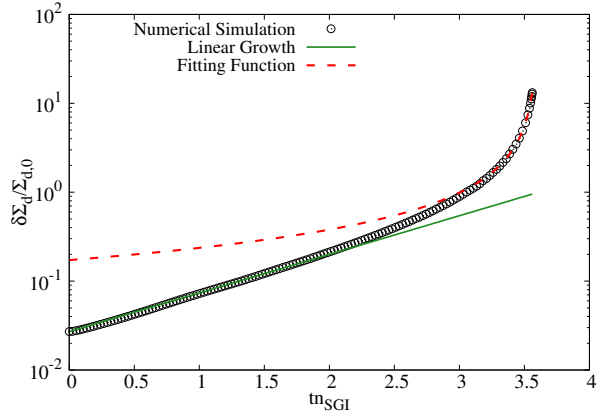}}
\end{center}
\caption{Results of the test calculation of the secular GI in the case of $t_{\tstop}\Omega=1$ and the fitting function of the non-linear growth. The horizontal axis is time normalized by the linear growth rate $n_{\mathrm{SGI}}$, and the vertical axis is the amplitude of the dust surface density perturbation normalized by the initial surface density at the center of the domain. The black circles represent the result of numerical calculation, the solid green line represents the linear growth, and the dashed red line shows the fitting function of the non-linear growth. In this case, we obtain $\tilde{t}_{\mathrm{c}}\simeq3.6$ and $q\simeq0.97$. \label{fig:fittaus1e0}}
\end{figure}
%\begin{figure*}[ht!]
%\centering
%\leavevmode
%%\includegraphics[bb = 0 0 288 216, width=0.45\hsize]{./f7a.pdf}
%%\includegraphics[bb = 0 0 288 216, width=0.45\hsize]{./f7a_wl.pdf}
%\includegraphics[bb = 0 0 288 216, width=7cm]{./f7a_wl.pdf}
%\hfil 
%%\includegraphics[bb = 0 0 288 216, width=0.45\hsize]{./f7b.pdf}
%%\includegraphics[bb = 0 0 288 216, width=0.45\hsize]{./f7b_wl.pdf}
%\includegraphics[bb = 0 0 288 216, width=7cm]{./f7b_wl.pdf}
%\caption{Dependence of the power $q$ (left panel) and the perturbed surface density $\delta_{\mathrm{c}}$ (right panel) on the stopping time. The horizontal axis represents the stopping time normalized by the angular velocity. \label{fig:deltac-taus}}
%\end{figure*}
%
%\begin{figure}[ht!]
%%\centering
%\begin{center}
%%\includegraphics[bb = 0 0 576 216, width=16cm]{./f7_dbl.pdf}
%%\FigureFile(8cm,16cm){f2.eps}
%\includegraphics[clip,width=8cm]{./f2.eps}
%\end{center}
%\caption{Dependence of the power $q$ (left panel) and the perturbed surface density $\delta_{\mathrm{c}}$ (right panel) on the stopping time. The horizontal axis represents the stopping time normalized by the angular velocity. \label{fig:deltac-taus}}
%\end{figure}
%
\begin{figure}[htbp]
%\hspace{-70pt}\raisebox{0pt}[0pt][0pt]{
	\begin{tabular}{c}
		\begin{minipage}{0.45\hsize}
			\begin{center}
				\hspace{-30pt}\raisebox{-225pt}[0pt][200pt]{\includegraphics[width=8cm]{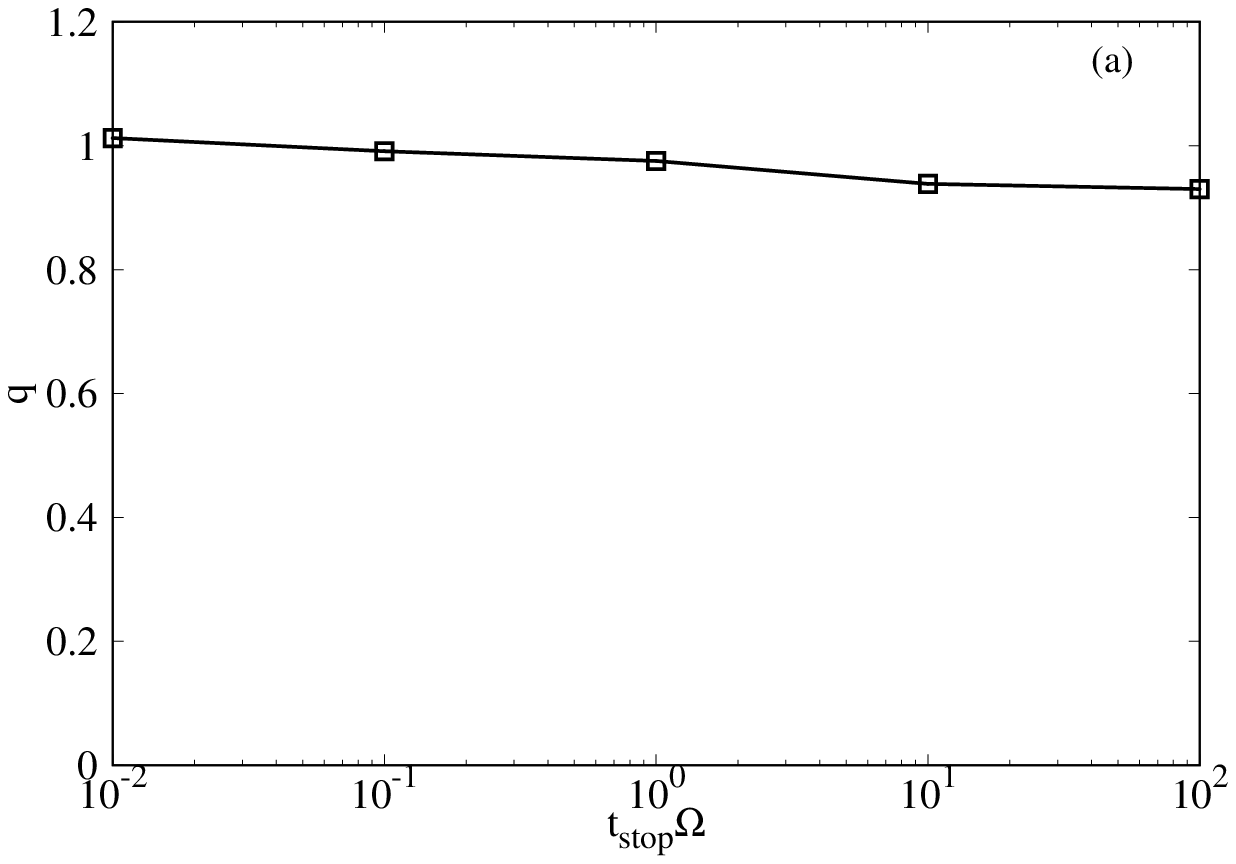}}
			\end{center}
		\end{minipage}
		\begin{minipage}{0.45\hsize}
			\begin{center}
				\hspace{-20pt}\raisebox{-225pt}[0pt][200pt]{\includegraphics[width=8cm]{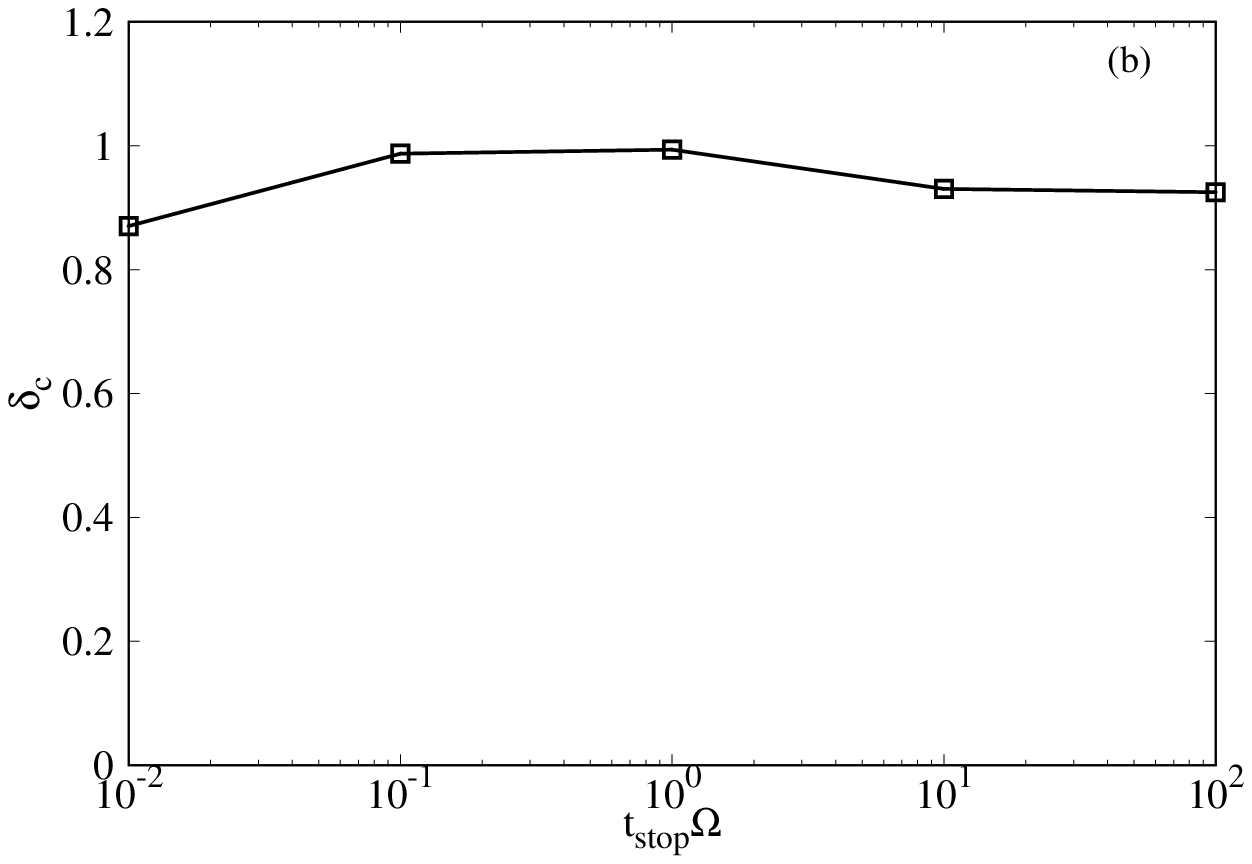}}
			\end{center}
		\end{minipage}
	\end{tabular}
%	}
\vspace{15pt}
\caption{Dependence of the power $q$ (left panel) and the perturbed surface density $\delta_{\mathrm{c}}$ (right panel) on the stopping time. The horizontal axis represents the stopping time normalized by the angular velocity. \label{fig:deltac-taus}}
\end{figure}
We also find that the power $q$ is almost unity for all cases. We can understand the reason as follows. If the dust surface density grows by more than the factor of $Q_{\dst}$, where $Q_{\dst}$ denotes the initial Toomre's $Q$ value for dust fluid, dust becomes self-gravitationally unstable. Since $Q_{\dst}=3$ in these simulations, we can consider dust to be self-gravitationally unstable in the non-linear stage of the secular GI. The time scale of the self-gravitational collapse is of the order of the free-fall time $t_{\mathrm{ff}}\sim 1/\sqrt{G\rho_{\dst}}$, where $\rho_{\dst}$ is the dust density. Supposed that the dust surface density $\Sigma_{\dst}$ is given by a product of the dust density and the dust's Jeans length $\lambda_{\mathrm{J}}\sim c_{\dst}/\sqrt{G\rho_{\dst}}$, we obtain $t_{\mathrm{ff}}\sim c_{\dst}/G\Sigma_{\dst}$. Therefore, the dust surface density is inversely proportional to the time $\tilde{t}-\tilde{t}_{\mathrm{c}}$. 
\subsection{Setups of simulation for wider radial extent}\label{sec:setup}
In this section, we show the results of the non-linear simulation. We first summarize the setup of our simulations. We use the piecewise polytropic equation of state with $c_{\s,0}\simeq 186$ m $\mathrm{s}^{-1}$   (equation \ref{eos}).
We assume that the size of dust grains is 3 mm and an intrinsic density is 3 $\mathrm{g}\; \mathrm{cm}^{-3}$. When we calculate $t_{\stp}$,  we use the Epstein drag law by assuming $\Sigma =\rho_{\mathrm{g}}/\sqrt{2\pi}H$, where $\rho_{\mathrm{g}}$ is gas density and $H\equiv c_{\mathrm{s}}/\Omega$, and we obtain $t_{\tstop}\Omega\simeq0.23$ in this case. For simplicity, we do not consider the dust size growth. The non-linear growth of the secular GI with the dust size growth is our future work. We consider a self-gravitationally stable disk around 1$\Msol$ star, in which initial surface densities of gas and dust, $\Sigma_{0}$ and $\Sigma_{\dst,0}$, are constant. Since pressure gradient force vanishes in such a disk, both gas and dust initially rotate with the Keplerian velocity. We assume that initial dust-to-gas mass ratio is equal to 0.1 and $c_{\dst}/c_{\s}$ is also equal to 0.1. The purpose for our choice of a somewhat unconventional disk model is to clearly understand the non-linear outcome of the secular GI under the condition where we can solve the dynamics very accurately. The initial values of the surface densities of gas and dust are set so that the Toomre parameters of both fluids are equal to 3 at $r=100$ au.
The radii of inner and outer boundaries, $r_{\mathrm{in}}$ and $r_{\mathrm{out}}$, are set to be 60 au and 140 au respectively. We adopt the fixed boundary condition. The number of cells is 2048, with which we divide the domain into cells with the same width. Time interval $\Delta t $ is determined with the way described in section \ref{sec:complinearana}. We initially adopt the following value as a softening length $h_{i+1/2}$:
\begin{equation}
h_{i+1/2}=\frac{r_{i+3/2}-r_{i-1/2}}{2} .\label{soft}
\end{equation}
%初期のメッシュ境界の位置を$r_{0,i+1/2}$とした時，$r_{i+3/2}-r_{i-1/2}>r_{0,i+3/2}-r_{0,i-1/2}$を満たす場合は式(\ref{soft})を用いて$h$を再評価する．このように再評価することで，解像度の悪い低密度領域からの重力を弱めた．
%初期の摂動は乱数を用いて与え，変位の振幅はメッシュ幅の0.1\%，ガスとダストの速度擾乱の振幅はそれぞれ$10^{-3}c_{\s,0},\;10^{-3}c_{\dst}$とした．\\
%----------------------------------
In reality, the non-linear growth of the secular GI may saturate since the self-gravity is weakened due to the effect of a disk's thickness \citep{Vandervoort1970, Shu1984}. To determine how the disk thickness changes in time and when the growth saturates, we need to conduct a multidimensional simulation. In this study, we use the softening length equation (\ref{soft}), which is smaller than the dust scale height $c_{\dst}/\Omega$, and investigate the evolution of the dust ring whose surface density saturates by the effect of softening the self-gravity. If $r_{i+3/2}-r_{i-1/2}>r_{0,i+3/2}-r_{0,i-1/2}$, where $r_{0,i+1/2}$ denotes the initial radius of a cell boundary, we recalculate $h$ using equation (\ref{soft}). By doing this re-evaluation, we weaken gravity from the low density region where the resolution is lower than the mean resolution. 

As the initial condition of numerical simulation, we put a random perturbation. The amplitude of displacement is 0.1\% of the mean width of cell, and those of velocity perturbation of gas and dust are given by $10^{-3}c_{\s,0}$ and $10^{-3}c_{\dst}$ respectively.

%------------3.3-----------------------------
\subsection{Overview of non-linear evolution of the secular GI}\label{sec:nonlinear}
\begin{figure}[ht!]
%\centering
\begin{center}
%\includegraphics[bb=0 0 288 216,clip,width=\hsize]{./f8.pdf}
%\includegraphics[bb=0 0 288 216,clip,width=8cm]{./f8.pdf}
%\FigureFile(8cm,8cm){f2.eps}
\hspace{-65pt}\raisebox{-225pt}[0pt][200pt]{\includegraphics[width=8cm]{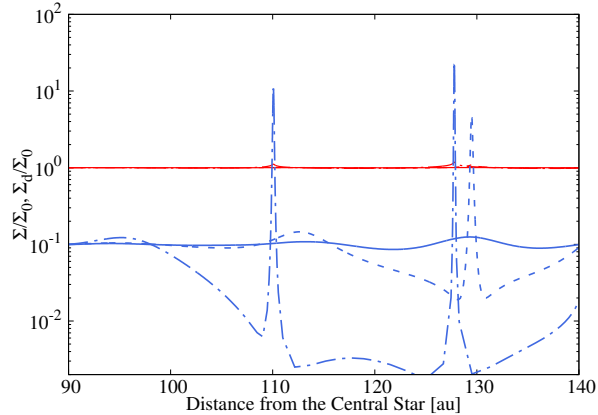}}
\end{center}
\caption{Snapshot of the distribution of the surface densities of dust and gas. The thick blue line shows the surface density of dust, the thin red line shows that of gas. Both of them are normalized by the unperturbed surface density of gas. The solid, dashed and dotted-dashed lines show the surface density distribution at $t=26148$ yr, $36479$ yr, $54689$ yr respectively. The most unstable mode of the secular GI at $r=110$ and $130$ au grows. While the gas surface density does not almost increase, the dust surface density becomes a hundred times larger than the initial surface density. \label{fig:rhoevolv}}
\end{figure}

\begin{figure}[ht!]
%\centering
\begin{center}
%\includegraphics[bb=0 0 288 216,clip,width=\hsize]{./f9.pdf}
%\includegraphics[bb=0 0 288 216,clip,width=8cm]{./f9.pdf}
%\FigureFile(8cm,8cm){f2.eps}
\hspace{-65pt}\raisebox{-225pt}[0pt][200pt]{\includegraphics[width=8cm]{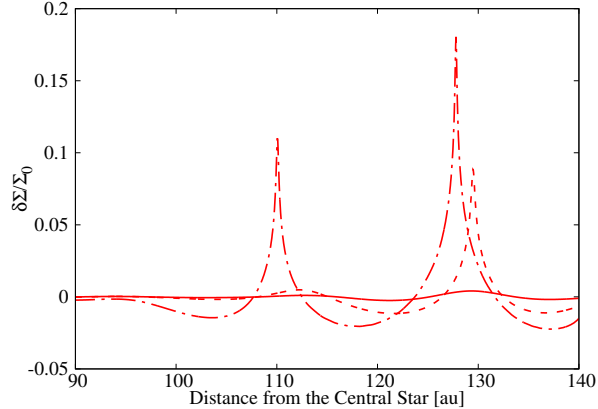}}
\end{center}
\caption{Snapshot of the distribution of the perturbed surface density of gas normalized by the unperturbed surface density. The solid, dashed and dotted-dashed lines show the surface density distribution at $t=26148$ yr, $36479$ yr, $54689$ yr respectively. Because of the stabilization by the pressure gradient force, the surface density increases up to only around 18 percent of the unperturbed surface density.\label{fig:drhogasevolv}}
\end{figure}

Figures {\ref{fig:rhoevolv}} and \ref{fig:drhogasevolv} are snapshots of surface density distributions of dust and gas. The most unstable mode of the secular GI grows at $r=110$ and $130$ au. The dust-to-gas ratio increases to the order of ten. While the surface density of gas increases up to only around 18 percent of the unperturbed surface density of gas, that of dust increases about hundredfold because the pressure gradient of dust is very small. 
We find that a width of the dust ring is very small, which we can understand as follows. Because the self-gravity of a very thin ring has same dependence on the shortest distance to the ring as that of a filament, we can write the self-gravity with a ring width $\Delta R$ and a line mass $M_{\mathrm{L}}$ as $\sim G M_{\mathrm{L}}/\Delta R$. On the other hand, the pressure gradient force can be written as $\sim c_{\dst}^2\Delta \Sigma_{\dst}/\Sigma_{\dst}\Delta R$, where $\Delta \Sigma_{\dst}$ is a change of the dust surface density in the ring. Because $\Delta \Sigma_{\dst}\sim\Sigma_{\dst}$ in the dust ring and $c_{\dst}$ is constant in this study, the dependence of the self-gravity on $\Delta R$ is same as that of the pressure gradient force. Thus, once dust become unstable and its density grows due to the self-gravity, the pressure gradient force can not suppress the growth, and an infinitesimally thin ring forms. In this simulation, the growth of the dust surface density is saturated because we calculate the self-gravity with the softening length. 
In such a dust concentrated ring, the planetesimal formation is expected to occur.
Therefore, the non-linear growth of the secular GI provides a powerful mechanism for the planetesimal formation.
%\begin{figure}[ht!]
%\centering
%%\includegraphics[bb=0 0 288 216,clip,width=\hsize]{./f8.pdf}
%\includegraphics[bb=0 0 288 216,clip,width=8cm]{./f8.pdf}
%\caption{Snapshot of the distribution of the surface densities of dust and gas. The thick blue line shows the surface density of dust, the thin red line shows that of gas. Both of them are normalized by the unperturbed surface density of gas. The solid, dashed and dotted-dashed lines show the surface density distribution at $t=26148$ yr, $36479$ yr, $54689$ yr respectively. The most unstable mode of the secular GI at $r=110$ and $130$ au grows. While the gas surface density does not almost increase, the dust surface density becomes a hundred times larger than the initial surface density. \label{fig:rhoevolv}}
%\end{figure}
%\begin{figure}[ht!]
%\centering
%%\includegraphics[bb=0 0 288 216,clip,width=\hsize]{./f9.pdf}
%\includegraphics[bb=0 0 288 216,clip,width=8cm]{./f9.pdf}
%\caption{Snapshot of the distribution of the perturbed surface density of gas normalized by the unperturbed surface density. The solid, dashed and dotted-dashed lines show the surface density distribution at $t=26148$ yr, $36479$ yr, $54689$ yr respectively. Because of the stabilization by the pressure gradient force, the surface density increases up to only around 18 percent of the unperturbed surface density.\label{fig:drhogasevolv}}
%\end{figure}
\begin{figure}[ht!]
%\centering
\begin{center}
%\includegraphics[bb=0 0 288 216,clip,width=\hsize]{./f10.pdf}
%\includegraphics[bb=0 0 288 216,clip,width=8cm]{./f10.pdf}
%\FigureFile(8cm,8cm){f2.eps}
\hspace{-65pt}\raisebox{-225pt}[0pt][200pt]{\includegraphics[width=8cm]{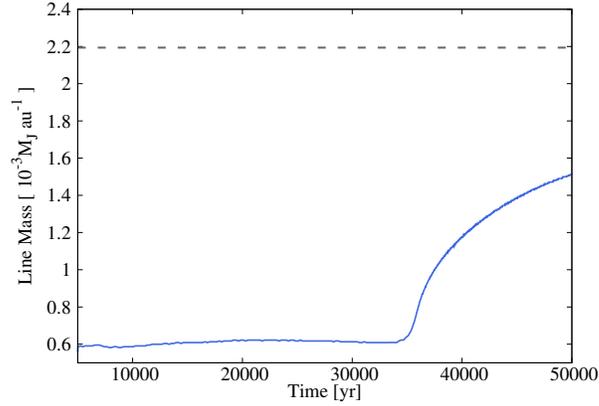}}
\end{center}
\caption{Time evolution of the line mass of the dust ring. The solid line shows the line mass of the dust ring, and the dashed line shows the critical line mass of filament in which a sound velocity is equal to $c_{\dst}$ we used in this work. We find that the line mass of the dust ring formed by the secular GI is comparable to the critical line mass. \label{fig:linemass}}
\end{figure}
Figure \ref{fig:linemass} shows the time evolution of a line mass of the dust ring formed by the secular GI. We define the width of the ring as a full width at half maximum of the dust surface density. After the secular GI grows with the constant line mass, dust around at a gap structure accretes onto the ring and the line mass increases. Figure \ref{fig:linemass} also shows that the line mass of the dust ring is comparable to the critical line mass of a filament $2c_{\dst}^2/G$. This feature comes from the similarity between the self-gravity of the ring and that of filament, whose mass that hold equilibrium by the pressure gradient force is determined by the critical line mass.  Gap structure also forms around the ring. 
\begin{figure}[ht!]
%\centering
\begin{center}
%\includegraphics[bb=0 0 288 216,clip,width=\hsize]{./f11.pdf}
%\includegraphics[bb=0 0 288 216,clip,width=8cm]{./f11.pdf}
%\FigureFile(8cm,8cm){f2.eps}
\hspace{-65pt}\raisebox{-225pt}[0pt][200pt]{\includegraphics[width=8cm]{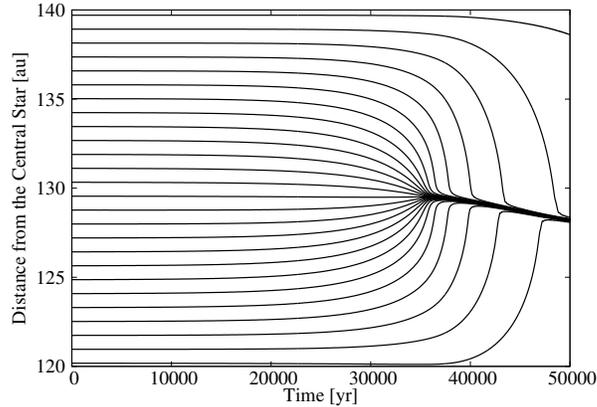}}
\end{center}
\caption{Time evolution of positions of dust cells. We reduce the number of cells in the plot. Because we solve the evolution with the Lagrangian fluid equation, the more cells there are, the larger the surface density is. The radius where the cells concentrate represents the radius of the ring. We can see that the ring moves inward after it forms. \label{fig:ringmigrating}}
\end{figure}
The dust-to-gas ratio in these gaps decreases by a factor of about ten. We find the dust-to-gas ratio reaches the minimum value at the vicinity of the dense ring.
\begin{figure}[ht!]
%\centering
\begin{center}
%\includegraphics[bb=0 0 288 216,clip,width=\hsize]{./f12.pdf}
%\includegraphics[bb=0 0 288 216,clip,width=8cm]{./f12.pdf}
%\FigureFile(8cm,8cm){f2.eps}
\hspace{-65pt}\raisebox{-225pt}[0pt][200pt]{\includegraphics[width=8cm]{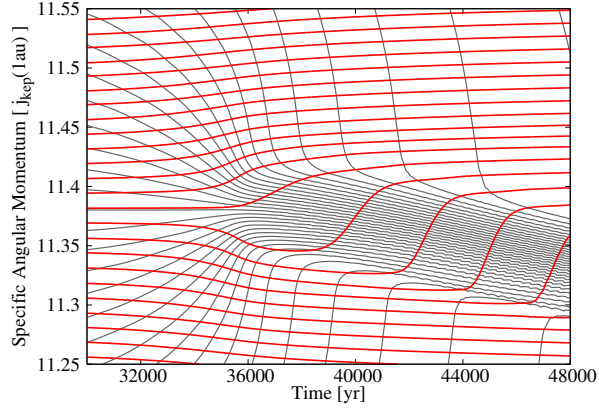}}
\end{center}
\caption{Time evolution of specific angular momentums of dust and gas. The thin gray line represents that of dust, the thick red line represent that of gas. The vertical axis shows the specific angular momentum normalized by the Keplerian specific angular momentum $j_{\mathrm{kep}}(r)$ at $r=$1au. We reduce the number of cells we plot here with a constant interval. The specific angular momentum of dust decreases with the dust ring migrating, while that of gas increases when the dust ring passes through. The high dust-to-gas ratio in the ring results in larger rate of change of specific angular momentum of gas than that of dust. \label{fig:ringmigrating-angmon}}
\end{figure}

We also find that the dust ring migrates inward. 
Because this disk initially has the flat density distribution, the radial drift of dust grains does not occur in the unperturbed state. Nevertheless, we find that non-linear growth of the secular GI results in a radial drift of the dust ring. 
Figure \ref{fig:ringmigrating} shows the time evolution of positions of dust cells. From $t\simeq 30000$yr the secular GI grows non-linearly, and the resulting dust ring moves inward slowly. In figure \ref{fig:ringmigrating-angmon} we plot the time evolution of the specific angular momentums of dust cells and gas cells around the dust ring.  Because both dust and gas rotate with almost Keplerian velocity, the time evolution of the specific angular momentum corresponds to that of the radius. We see dust loses its angular momentum and moves inward while gas gets angular momentum from dust.

%\newpage

\section{Discussions: semi-analytic model of ring migration}\label{sec:discussion}

We find that the migration of the ring is driven by the self-gravity of itself. In this section, we compare the velocity derived by using a semi-analytic model with the result of our numerical simulation.

\begin{figure}[ht!]
%\centering
\begin{center}
%\includegraphics[bb=0 0 288 216,clip,width=\hsize]{./f13.pdf}
%\includegraphics[bb=0 0 288 216,clip,width=8cm]{./f13.pdf}
%\FigureFile(8cm,8cm){f2.eps}
\hspace{-65pt}\raisebox{-225pt}[0pt][200pt]{\includegraphics[width=8cm]{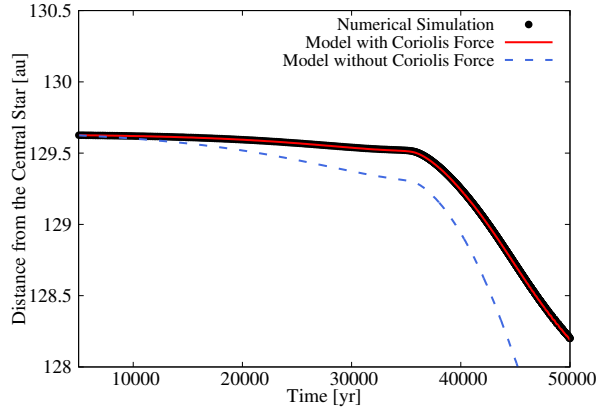}}
\end{center}
\caption{Comparison with the ring's radius (the black filled circles) and the radius calculated with the one-fluid model. The dashed blue line corresponds to the model without the Coriolis force, and the solid red line corresponds to that with the Coriolis force. We assume $t_{\tstop}\Omega_{\mathrm{kep}}=0.2338$ for the models. The solid red line has good agreement with the results of the numerical simulation. \label{fig:1f-model}}
\end{figure}

First we consider the dust motion. The specific angular momentum of gas increases when the dust ring passes through, and gas moves outward (figure \ref{fig:ringmigrating-angmon}). The increase, however, is only one percent of the initial value, and the specific angular momentum does not change after the dust ring passed. Therefore, we may neglect the time-dependence of gas profile in the analysis of dust ring migration, $u_r=0$. The equation of the radial motion of dust is 
\begin{equation}
\frac{dv_r}{dt} = \frac{v_{\theta}^2}{r}-\frac{G M_{\ast}}{r^2}-\frac{c_{\dst}^2}{\Sigma_{\dst}}\frac{\partial \Sigma_{\dst}}{\partial r}-\frac{\partial \Phi}{\partial r}-\frac{v_{r}}{t_{\tstop}}.
\end{equation}
Assuming that the difference between the orbital velocity and the Keplerian velocity is small and the radial velocity is given by the terminal velocity, we obtain
\begin{equation}
v_r=t_{\tstop}\left[2\delta v_{\theta}\Omega_{\mathrm{kep}} -\frac{c_{\dst}^2}{\Sigma_{\dst}}\frac{\partial \Sigma_{\dst}}{\partial r}-\frac{\partial \Phi}{\partial r}\right],\label{dust_model}
\end{equation}
where $\Omega_{\mathrm{kep}}\equiv \sqrt{G M_{\ast}/r^3}$. The first term on the right hand side corresponds to the Coriolis force. We evaluate the right hand side of equation (\ref{dust_model}) at the cell whose radius traces the ring's center. To compare the numerical result with the model, we integrate in time the velocity derived from the model, and evaluate the time evolution of the ring's radius. We compare the model with simulation from $t\simeq5000$ yr when the surface density perturbation grows to about $4\times10^{-3}$ in order to trace the most unstable mode. In figure \ref{fig:1f-model}, we compare the result of numerical simulation with the model without the Coriolis force and the model with it. Although the Coriolis force decelerates the migrating ring, the velocity is basically determined by the sum of the pressure gradient force and the self-gravity. Thus, we may explain the origin of the dust ring migration by neglecting the Coriolis force as in the following. First, we decompose the radial component of  the ring's self-gravity into two parts: a component of which inner and outer parts are symmetric with respect to the center of the ring and a homogeneous component as in equation (\ref{divide-ring}). 
The symmetric component corresponds to a force that makes the ring's width small. 
\begin{figure}[ht!]
\centering
%\includegraphics[bb=0 0 288 216,clip,width=\hsize]{./f14.pdf}
%\includegraphics[bb=0 0 288 216,clip,width=8cm]{./f14.pdf}
%\includegraphics[clip,width=8cm]{./f14.eps}
%\FigureFile(8cm,8cm){f2.eps}
\hspace{-65pt}\raisebox{-225pt}[0pt][200pt]{\includegraphics[width=8cm]{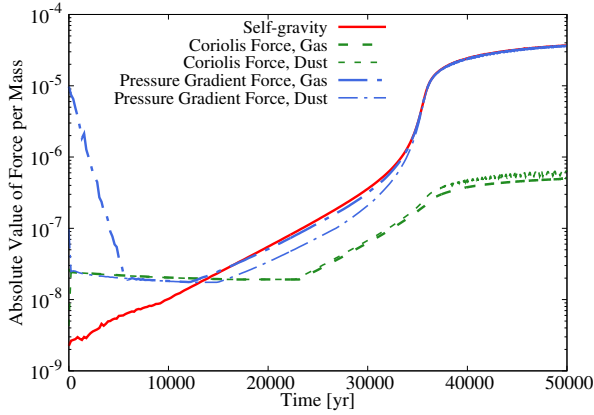}}
\caption{Time evolution of the maximums of the absolute values of the forces normalized by $GM_{\ast}/R_0^2$, where $R_0$ is equal to 1 au. The solid red line represents the self-gravity, the dotted-dashed blue lines show the pressure gradient force, and the dashed green lines show the sum of the curvature term and the central star gravity. The thin lines represent the forces exerted on dust, and the thick lines show that exerted on gas. The self-gravity almost balances with the gas pressure gradient force. From $t\simeq 35000$yr when the dust ring starts to migrate, the dust pressure gradient force is comparable to the self-gravity.\label{fig:force-amplitude}}
\end{figure}
The homogeneous component corresponds to a force that drives the dust ring migration. 
\begin{equation}
-\frac{\partial \Phi}{\partial r}=-\frac{\partial \Phi_{\mathrm{sym}}}{\partial r}-\frac{\partial \Phi_{\mathrm{h}}}{\partial r}\label{divide-ring}
\end{equation}
When we consider the growth of the secular GI, the self-gravity dominates the other forces acting on dust. The growth stops when the symmetric component of the self-gravity is comparable to the pressure gradient force of dust. Because the ring migrates after the growth saturates, we can consider that the self-gravity balances with dust's pressure gradient force (see figure \ref{fig:force-amplitude}). Therefore, we can write the migrating velocity as follows:
\begin{equation}
v_r\simeq -t_{\tstop}\frac{\partial \Phi_{\mathrm{h}}}{\partial r}\label{dust-model-onlySG}
\end{equation}
The velocities at the time $t=40000$ yr calculated with equation (\ref{dust_model}) is about $-0.8 \times 10^{-4}$ au $\mathrm{yr}^{-1}$, which is in good agreement with the right hand side of equation (\ref{dust-model-onlySG}) ($v_r\sim-1.0\times 10^{-4}$au $\mathrm{yr}^{-1}$). We can conclude that the ring migrates with the terminal velocity determined by the friction and the homogeneous component of the self-gravity.
%\begin{figure}[ht!]
%\centering
%\includegraphics[bb=0 0 288 216,clip,width=\hsize]{./f13.pdf}
%\caption{Comparison with the ring's radius (the black filled circles) and the radius calculated with the one-fluid model. The dashed yellow line corresponds to the model without the Coriolis force, and the solid orange line corresponds to that with the Coriolis force. We assume $t_{\tstop}\Omega_{\mathrm{kep}}=0.2338$ for the models. The orange line has good agreement with the results of the numerical simulation. \label{fig:1f-model}}
%\end{figure}
%\begin{figure}[ht!]
%\centering
%\includegraphics[bb=0 0 288 216,clip,width=\hsize]{./f14.pdf}
%\caption{Time evolution of the maximums of the absolute values of the forces. The solid black line represents the self-gravity, the dotted-dashed lines show the pressure gradient force, and the dashed lines show the sum of the curvature term and the central star gravity. The black lines represent the forces exerted on dust, and the gray lines show that exerted on gas. The self-gravity almost balances with the gas pressure gradient force. From $t\simeq 35000$yr when the dust ring starts to migrate, the dust pressure gradient force is comparable to the self-gravity.\label{fig:force-amplitude}}
%\end{figure}

\begin{figure}[ht!]
%\centering
\begin{center}
%\includegraphics[bb=0 0 288 216,clip,width=\hsize]{./f15.pdf}
%\includegraphics[bb=0 0 288 216,clip,width=8cm]{./f15.pdf}
%\FigureFile(8cm,8cm){f2.eps}
\hspace{-65pt}\raisebox{-225pt}[0pt][200pt]{\includegraphics[width=8cm]{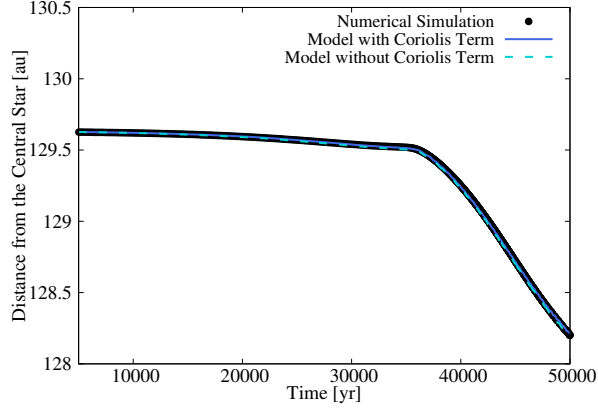}}
\end{center}
\caption{Comparison with the ring's radius (the black filled circles) and the radius calculated with the two-fluids model. The dashed sky blue line corresponds to the model without the first term on the right hand side of equation (\ref{gas-dust-model}), which corresponds to the Coriolis force. The solid blue line corresponds to the model with it. We also assume $t_{\tstop}\Omega_{\mathrm{kep}}=0.2338$. The difference between both lines is small, which means the effect of the relative orbital velocities is small. \label{fig:2f-model}}
\end{figure}

We also compare the model, in which we consider both gas and dust with the numerical simulation. We assume that the radial velocity of gas is zero again and the difference between the orbital velocity and the Keplerian velocity $\delta u_{\theta}$ is small. Then we can write the equation of the radial motion of gas as:
\begin{equation}
2\delta u_{\theta}\Omega_{\mathrm{kep}} -\frac{c_{\s}^2}{\Sigma}\frac{\partial \Sigma}{\partial r}-\frac{\partial \Phi}{\partial r}+\frac{\Sigma_{\dst}}{\Sigma}\frac{v_{r}}{t_{\tstop}}=0.\label{gas_model_eom}
\end{equation}
With using equations (\ref{dust_model}) and (\ref{gas_model_eom}) and eliminating the self-gravity, we obtain
\begin{equation}
v_r=\frac{t_{\tstop}}{1+\Sigma_{\dst}/\Sigma}\left[2(\delta v_{\theta}-\delta u_{\theta})\Omega_{\mathrm{kep}} +\frac{c_{\s}^2}{\Sigma}\frac{\partial \Sigma}{\partial r}-\frac{c_{\dst}^2}{\Sigma_{\dst}}\frac{\partial \Sigma_{\dst}}{\partial r}\right].\label{gas-dust-model}
\end{equation}
In figure \ref{fig:2f-model}, we compare the ring's radius obtained by the numerical simulation with the radius calculated with the two-fluids model. We find the model has good agreement with the numerical result, and the effect of the first term on the right hand side of equation (\ref{gas-dust-model}) is small. 
We can understand the reason why the term that comes from the relative velocity in the azimuthal direction is small as follows. Since the migration timescale is sufficiently longer than the Coriolis timescale, we can apply the terminal velocity approximation in the azimuthal direction. The equation of motion of dust in the azimuthal direction is  
\begin{equation}
-\frac{v_{\theta}v_r}{r}-\frac{v_{\theta}-u_{\theta}}{t_{\tstop}}=0.
\end{equation}
This can be written as follows:
\begin{equation}
|v_{\theta}-u_{\theta}|=t_{\tstop}\Omega|v_r|.
\end{equation}
%From this equation, we find the first term in equation (\ref{gas-dust-model}) is $\sim(t_{\tstop}\Omega)^2(1+\Sigma_{d}/\Sigma)^{-1}v_r$. In this simulation $t_{\tstop}\Omega\simeq0.23$, and then we obtain $(t_{\tstop}\Omega)^2(1+\Sigma_{d}/\Sigma)^{-1}|v_r|\ll |v_r|$.
%\begin{figure}[ht!]
%\centering
%%\includegraphics[bb=0 0 288 216,clip,width=\hsize]{./f15.pdf}
%\includegraphics[bb=0 0 288 216,clip,width=8cm]{./f15.pdf}
%\caption{Comparison with the ring's radius (the black filled circles) and the radius calculated with the two-fluids model. The dashed sky blue line corresponds to the model without the first term on the right hand side of equation (\ref{gas-dust-model}), which corresponds to the Coriolis force. The solid blue line corresponds to the model with it. We also assume $t_{\tstop}\Omega_{\mathrm{kep}}=0.2338$. The difference between both lines is small, which means the effect of the relative orbital velocities is small. \label{fig:2f-model}}
%\end{figure}

In this simulation, the migration velocity of the dust ring is about $10^{-5}-10^{-4}$ au $\mathrm{yr}^{-1}$. From the radius $r=130$ au where the the secular GI likely grows,  it takes $10^{6}$yr for the ring to reach the center of the disk, which is of the order of a life-time of a disk. Thus, we conclude the migrating speed is very slow.

Finally, we compare the migrating velocity of the dust ring and the radial drift of dust grains that is thought to occur in disks with non-uniform gas pressure profile. In this study, we have only focused on the simplified case where the radial profile of the unperturbed surface densities of dust and gas are uniform. In general, however, there is a surface density gradient in a protoplanetary disk, which results in relative velocity between dust and gas in the orbital direction and consequent radial drift of dust grains \citep{Nakagawa1986}.  The drift velocity $v_{\mathrm{drift}}$ is written as follows:
\begin{equation}
v_{\mathrm{drift}}=-2\eta r\Omega_{\mathrm{kep}}\frac{t_{\tstop}\Omega_{\mathrm{kep}}}{\left(1+\epsilon\right)^2+\left(t_{\tstop}\Omega_{\mathrm{kep}}\right)^2},
\end{equation}
where
\begin{equation}
\eta\equiv -\frac{1}{2}\left(\frac{c_{\s}}{r\Omega_{\mathrm{kep}}}\right)^2\frac{d\ln P}{d\ln r},
\end{equation}
and $\epsilon$ is the dust-to-gas density ratio. Assuming the minimum-mass solar nebula \citep{Hayashi1981},
\begin{equation}
\eta\simeq 1.8\times 10^{-3}\left(\frac{r}{1 \mathrm{au}}\right)^{1/2},
\end{equation}
\begin{equation}
v_{\mathrm{drift}}\simeq -2.2\times 10^{-2}\frac{t_{\tstop}\Omega_{\mathrm{kep}}}{\left(1+\epsilon\right)^2+\left(t_{\tstop}\Omega_{\mathrm{kep}}\right)^2} \;\;\mathrm{au}\; \mathrm{yr}^{-1}.\label{drift_vel}
\end{equation}
Assuming $\epsilon = 0.1$ and $t_{\tstop}\Omega_{\mathrm{kep}}=0.23$, which are the initial value in our simulation, we obtain $v_{\mathrm{drift}}\simeq -4.1\times 10^{-3}\; \mathrm{au}\; \mathrm{yr}^{-1}$, which is faster than the migrating velocity of the dust ring. On the other hand, we may expect that the migrating velocity of the dust ring is about $10^{-5}-10^{-4}\; \mathrm{au}\; \mathrm{yr}^{-1}$ as we obtained in this work even in a disk with non-uniform gas pressure profile because the velocity is determined by the self-gravity of the ring formed through the non-linear growth of the secular GI. Thus, we see that the migrating velocity of the dust ring is much slower than the radial drift velocity of a dust grain in the unperturbed disk. 

\section{Conclusion}\label{sec:conclusion}
In this work, we investigate the non-linear growth of the secular GI using the numerical simulation with new hydrodynamic symplectic scheme. Applying the symplectic method to the numerical hydrodynamics, we can realize a long term calculation without numerical dissipation due to advection. With our new scheme, we conduct the non-linear simulation of the secular GI in a Keplerian rotating disk with the constant surface density. From the results, we find the followings: 
\begin{enumerate}
\item the maximum surface density of dust becomes at least hundreds times larger than the initial value, while that of gas increases only by around 18\% of the unperturbed surface density. 
%As a result, growth of the secular GI results in  the formation of the ring with the dust-to-gas mass ratio $\sim$10.
As a result, the ring with the dust-to-gas mass ratio $\sim$10 forms through the growth of the secular GI.
 In such a dust-dominated ring, the planetesimal formation is expected to occur. This results indicates that the non-linear growth of the secular GI provides a powerful mechanism for the planetesimal formation.
\item The line mass of the dust ring is about the critical line mass $2c_{\dst}^2/G$. Thus, the observation of the dust ring with the critical line mass indicates that it is formed by the secular GI.
\item we find that the dust ring created by the non-linear growth of the secular GI migrates inward. We semi-analytically derive the migration speed from the equation of motion with the terminal velocity approximation (equations (\ref{dust_model}) and (\ref{gas-dust-model})). Our model has good agreement with the numerical results (figures \ref{fig:1f-model} and \ref{fig:2f-model}).
\end{enumerate}

As mentioned above, we adopt the constant surface density distribution as the unperturbed state to compare the non-linear evolution of the secular GI with the linear stability analysis in this work. Our future work focuses on the non-linear growth of the secular GI in a disk with the surface density gradient where dust initially drifts inward due to the relative velocity in the orbital direction. We need to investigate how the instability condition is modified by the effect of the radial drift of dust grains due to the possible pressure gradient of the disk. If the dust grains concentrate in rings through the growth of the secular GI even with the radial drift of the dust grain, the drift velocity of the dust grains in the rings decreases (equation (\ref{drift_vel})), which may be a solution for the radial drift problem and result in the planetesimal formation in the rings. We need to confirm this scenario with numerical simulations in our future work. In addition, a turbulent diffusion should be considered to understand the non-linear growth of the secular GI. Since the turbulent diffusion strongly stabilizes the secular GI, it may determine the saturation of the growth of the secular GI. Moreover we expect that the turbulence widens the width of rings formed through the non-linear growth of the secular GI. This effect is expected to work favorably in explaining the width of the observed ring as a non-linear outcome of the secular GI. We also consider a dust growth that must occur during the timescale of growth of the secular GI that is much longer than the Keplerian orbital period in our future work.

\bigskip
\begin{ack}
%Acknowledgement should be placed at end of main text.
%(NOT after the Appendix.)
This work was supported by JSPS KAKENHI Grant Number 16H02160 and 23244027.
\end{ack}

\appendix 
\section{Symplectic integration of non-dissipative hydrodynamics}\label{sec:ap1}
In this appendix, we show a formulation of the symplectic method of non-dissipative hydrodynamics that we have newly developed. First, we consider one dimensional sound wave in the Cartesian coordinate, and show how to derive equations of motion of cell-boundaries and results of test calculations. Next, we derive the equation in cylindrical coordinate.
%\begin{figure*}[t!]
%\centering
%\leavevmode
%%\includegraphics[bb = 0 0 512 389, width=0.425\hsize]{./f16a.pdf}
%%\includegraphics[bb = 0 0 512 389, width=0.425\hsize]{./f16a_wl.pdf}
%\includegraphics[bb = 0 0 512 389, width=7cm]{./f16a_wl.pdf}
%\hfil 
%%\includegraphics[bb = 0 0 512 389, width=0.425\hsize]{./f16b.pdf}
%%\includegraphics[bb = 0 0 512 389, width=0.425\hsize]{./f16b_wl.pdf}
%\includegraphics[bb = 0 0 512 389, width=7cm]{./f16b_wl.pdf}
%\caption{Results of the test calculation of a propagation of linear sound wave. The left figure shows a time variation of the error in the total energy．The amplitude does not increase monotonically even in the case of a calculation for the longer time. The right figure shows a density profile propagating along the x direction. The solid green lines show the exact solution, while the black circles show the numerical simulation. Although there is dispersive error, the amplitude keeps constant over 100 periods. \label{fig:ap1f1}}
%\end{figure*}

We derive the time-evolution equation from the action principle. The Lagrangian of a plane wave is given by the following:
\begin{equation}
L=\int dx\left[\rho\left(\frac{\dot{x}^2}{2}-u \right)\right],
\end{equation}
where $\dot{x}$ represents a time derivative of position $x$, $\rho$ is the density, and $u$ is the specific internal energy. Assuming the barotropic relation $P=P(\rho)$, the specific internal energy is 
\begin{equation}
u=\int \frac{P}{\rho^2}d\rho.\label{internal_e}
\end{equation}
Next, we discretize the Lagrangian as the following: 
\begin{equation}
L=\sum_{i=1}^{N-1}\left(m_{i+1/2}\frac{\dot{x}_{i+1/2}^2}{2}\right)-\sum_{i=1}^{N}m_{i}u_i ,\label{discri-lag}
\end{equation}
\begin{equation}
m_{i+1/2}=\frac{m_{i+1}+m_i}{2},
\end{equation}
where $N$ denotes the number of cells, the index $i$ represents a physical property defined in the $i$-th cell. A position of a boundary between the $i$-th cell and the $(i+1)$-th cell is denoted by $x_{i+1/2}$. The mass of the $i$-th cell $m_{i}$ is constant in time. The density $\rho_{i}$ is given by
\begin{equation}
\rho_{i}=\frac{m_i}{x_{i+1/2}-x_{i-1/2}}.
\end{equation}
Substituting the equation (\ref{discri-lag}) into the Euler-Lagrange equation
\begin{equation}
\frac{d}{d t}\left( \frac{\partial L}{\partial \dot{x}_{i+1/2}}\right)-\frac{\partial L}{\partial x_{i+1/2}} = 0,
\end{equation}
we obtain
\begin{equation}
m_{i+1/2}\ddot{x}_{i+1/2}=-\frac{\partial}{\partial x_{i+1/2}}\sum_{j=1}^{N}m_{j}u_j.\label{pre_eq_lag}
\end{equation}
With the use of equation (\ref{internal_e}), the right hand side of equation (\ref{pre_eq_lag}) is
\begin{eqnarray}
-\frac{\partial}{\partial x_{i+1/2}}\sum_{j=1}^{N}m_{j}u_j &=&-\frac{\partial}{\partial x_{i+1/2}}(m_{i+1}u_{i+1}+m_{i}u_i),\nonumber \\
&=&-\frac{\partial \rho_{i+1}}{\partial x_{i+1/2}}\frac{\partial \left(m_{i+1}u_{i+1}\right)}{\partial \rho_{i+1}}\nonumber\\
&&-\frac{\partial \rho_{i}}{\partial x_{i+1/2}}\frac{\partial \left(m_{i}u_{i}\right)}{\partial \rho_{i}},\nonumber \\
&=&-P_{i+1}+P_{i}.
\end{eqnarray}
Finally, we obtain the following equation of motion of the cell boundary:
\begin{equation}
m_{i+1/2}\ddot{x}_{i+1/2}=-(P_{i+1}-P_{i}).\label{soundeom}
\end{equation}
We integrate equation (\ref{soundeom}) with the symplectic method. Equation (\ref{soundeom}) is second-order accurate in $\Delta x_i$, which we can prove as follows. The Taylor-series expansion of the right hand side of equation (\ref{soundeom}) around $x_{i+1/2}$ is 
\begin{equation}
m_{i+1/2}\ddot{x}_{i+1/2}=-\frac{\Delta x_{i}+\Delta x_{i+1}}{2}\frac{d P}{d x}+\mathcal{O}(\Delta x_{i}^3),
\end{equation}
where we define $x_{i}$ as $\left( x_{i+1/2}+x_{i-1/2}\right)/2$, and assume $\Delta x_{i+1}=\Delta x_i +\mathcal{O}(\Delta x_{i}^2)$. The mass $m_{i+1/2}$ can be written as follows:
\begin{eqnarray}
m_{i+1/2}&=&\frac{1}{2}\left(m_{i}+m_{i+1} \right)\nonumber\\
&=&\frac{1}{2}\left(\rho_{i}\Delta x_i+\rho_{i+1}\Delta x_{i+1} \right)\nonumber\\
&=&\frac{\rho_{i+1/2}}{2}\left( \Delta x_{i}+\Delta x_{i+1}\right)+\mathcal{O}(\Delta x_i^3).\label{m_expan}
\end{eqnarray}
Dividing the equation (\ref{soundeom}) by $m_{i+1/2}$, we obtain
\begin{equation}
\ddot{x}_{i+1/2}=-\frac{1}{\rho_{i+1/2}}\frac{d P}{d x}+\mathcal{O}(\Delta x_{i}^2).
\end{equation}

We test our method against the propagation of a sound wave in a periodic box with the length $L$. We use $N=$128 cells and set these widths to be equal.  We use the isothermal equation of state with the sound speed $c_{\s}$. A time-step $\Delta t$ is taken to be $0.5L/Nc_{\s}$. We adopt that $L$ and $c_{\s}$ are unity and the unperturbed density is 2 in code units. We initially set up an amplitude of displacement $\xi=1.0\times10^{-6}$ and a wavelength $\lambda=0.5$. We integrate equation (\ref{soundeom}) using the leap-frog method that is one of the symplectic integrator and the second-order scheme. 
\begin{figure}[htbp]
%\hspace{-70pt}\raisebox{0pt}[0pt][0pt]{
	\begin{tabular}{c}
		\begin{minipage}{0.45\hsize}
			\begin{center}
				%\hspace{-70pt}\raisebox{-69.5pt}[0cm][0cm]{\includegraphics[clip,width=7.38cm]{./f16a.eps}}
				%\hspace{-30pt}\raisebox{-211.5pt}[0pt][200pt]{\includegraphics[width=7.38cm]{./f16a.eps}}
				\hspace{-30pt}\raisebox{-211.0pt}[0pt][200pt]{\includegraphics[width=8cm]{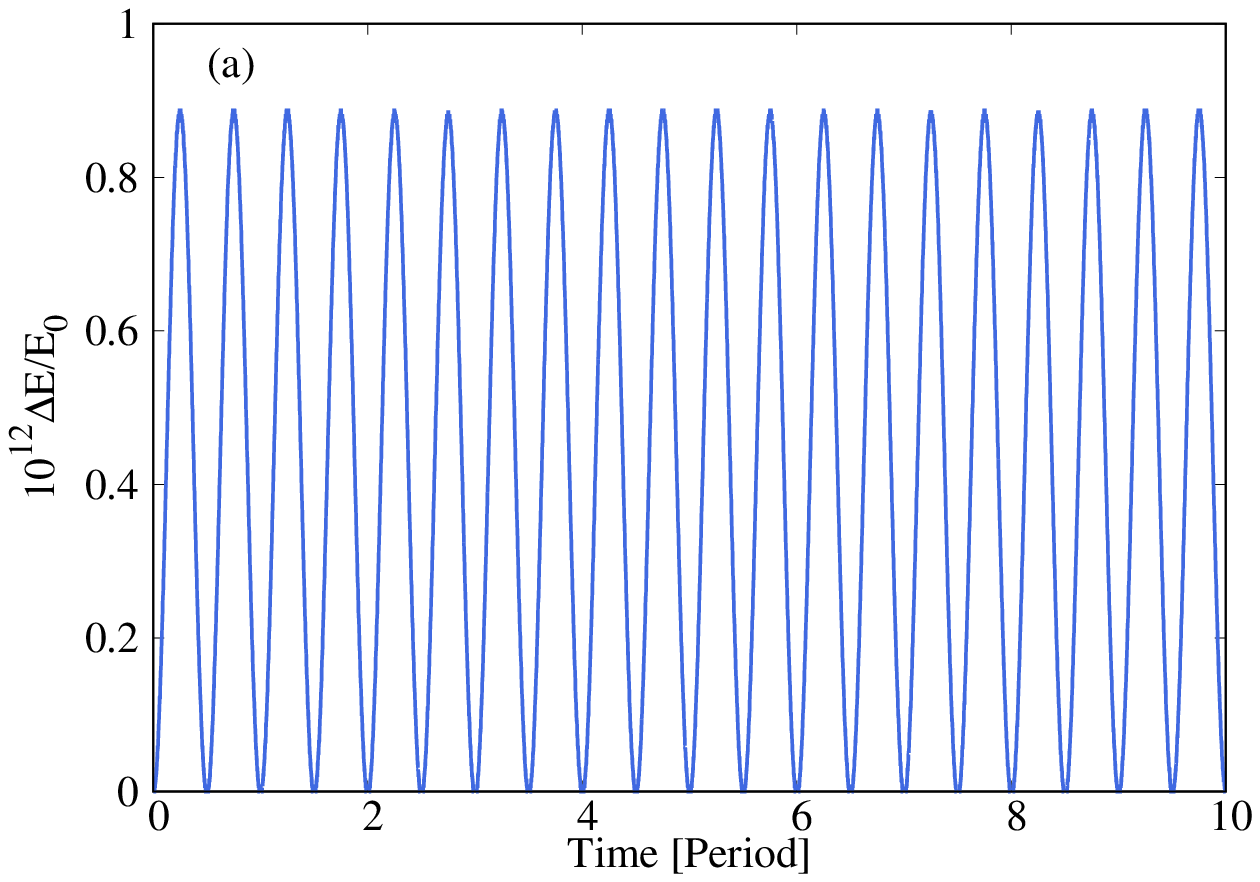}}
			\end{center}
		\end{minipage}
		\begin{minipage}{0.45\hsize}
			\begin{center}
				\hspace{-30pt}\raisebox{-225pt}[0pt][200pt]{\includegraphics[width=8cm]{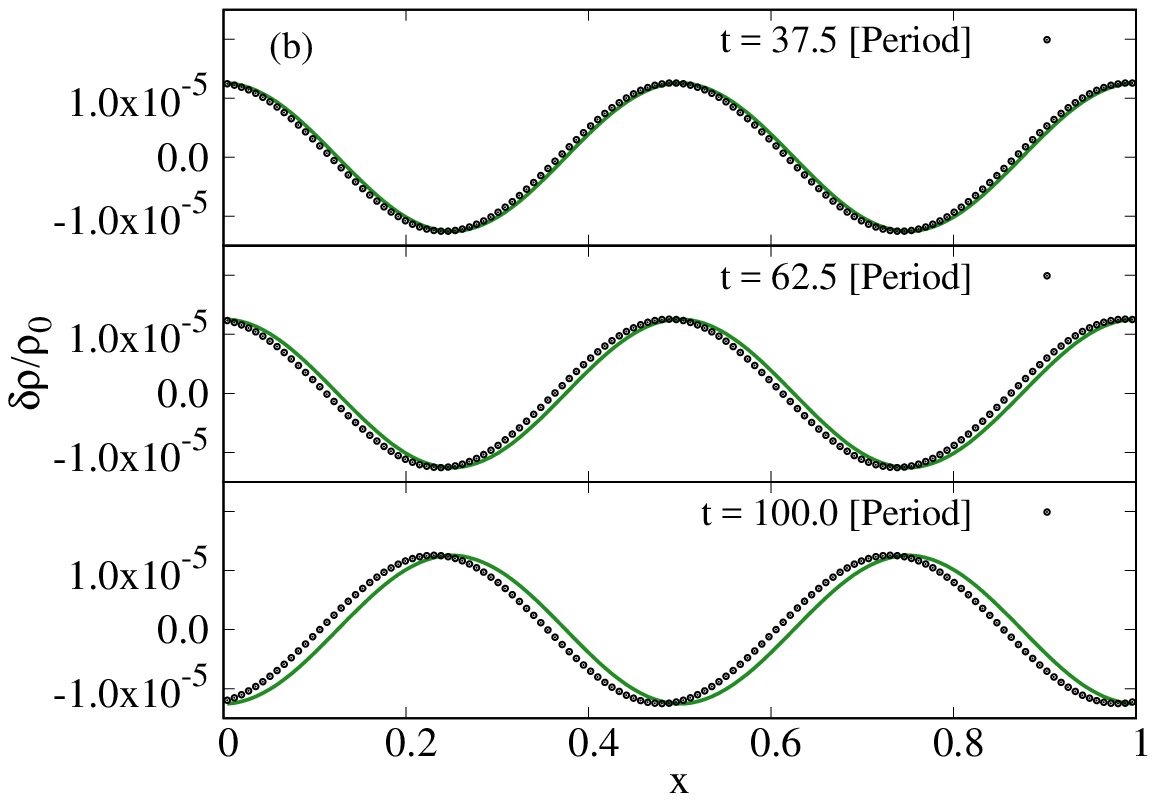}}
			\end{center}
		\end{minipage}
	\end{tabular}
%	}
\vspace{15pt}
\caption{Results of the test calculation of a propagation of linear sound wave. The left figure shows a time variation of the error in the total energy that is denoted by $\Delta E$. The initial total energy is represented by $E_0$. The amplitude does not increase monotonically even in the case of a calculation for the longer time. The right figure shows a density profile propagating along the x direction. The unperturbed density is denoted by $\rho_0$, and the amplitude of the perturbation of the density is denoted by $\delta\rho$. The solid green lines show the exact solution, while the black circles show the numerical simulation. Although there is dispersive error, the amplitude keeps constant over 100 periods. \label{fig:ap1f1}}
\end{figure}
Figure \ref{fig:ap1f1} shows the time variation of the error in the total energy and the time evolution of the density distribution. The oscillation of the error in the total energy shown in figure \ref{fig:ap1f1} is one of the feature of the symplectic integrator. Although there is the dispersive error in density distribution, the numerical dissipation does not occur. The amplitude of the sound wave keeps constant over 100 periods.

\begin{figure}[ht!]
%\centering
\begin{center}
%\includegraphics[bb = 0 0 512 389, width=\hsize]{./f17.pdf}
%\includegraphics[bb = 0 0 512 389, width=8cm]{./f17.pdf}
%\FigureFile(8cm,8cm){f2.eps}
\hspace{-65pt}\raisebox{-225pt}[0pt][200pt]{\includegraphics[width=8cm]{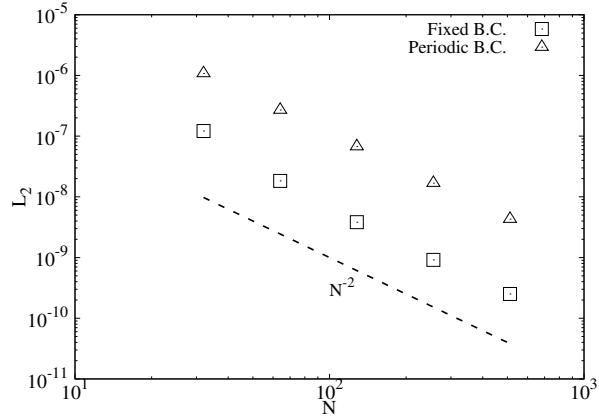}}
\end{center}
\caption{Results of the convergence test. This figure shows dependence of $L_2$ norm errors on the number of cells. The symbols $\Box$ and $\triangle$ represent the results in the cases of using the fixed boundary condition and the periodic boundary condition, respectively.\label{fig:ap1f2}}
\end{figure}

We conduct a convergence test to check the spatial accuracy of our scheme. The unperturbed density, the wavelength and the amplitude are the same as those of the above test. We conduct two test calculations. One is the propagating sound wave test using the periodic boundary. Another one is the standing wave test using the fixed boundary condition. We set $\Delta t=0.1 L/512c_{\s}$ in both tests and integrate until $t=3$ [periods]. Figure \ref{fig:ap1f2} shows the $N$ dependence of $L_2$ norm error of the density distribution at the last time step. The error is found to be proportional to $N^{-2}$, which represents our scheme has the second-order spatial accuracy.

In the case of an infinitesimally thin and axisymmetric disk, on which a potential force is exerted, we can formulate our symplectic scheme. In a cylindrical coordinate ($r,\theta$), we use the following discretized Lagrangian and the surface density defined at each cell:
\begin{equation}
L=\sum_{i=1}^{N-1}m_{i+1/2}\left(\frac{\dot{\vec{r}}^2_{i+1/2}}{2}+\Phi(r_{i+1/2}) \right)-\sum_{i=1}^Nm_{i}u_i,\label{cyl-lag}
\end{equation}
\begin{equation}
\dot{\vec{r}}^2_{i+1/2}\equiv \dot{r}_{i+1/2}^2+r^2_{i+1/2}\dot{\theta}_{i+1/2}^2,
\end{equation}
\begin{equation}
\Sigma_i=\frac{m_i}{\pi\left(r_{i+1/2}^2-r_{i-1/2}^2 \right)},
\end{equation}
where $\Phi$ is the potential. Substituting the equation (\ref{cyl-lag}) into the Euler-Lagrange equation, we obtain 
\begin{eqnarray}
m_{i+1/2}\ddot{r}_{i+1/2}&=&\frac{J_{i+1/2}^2}{m_{i+1/2}r_{i+1/2}^3}-2\pi r_{i+1/2}(P_{i+1}-P_{i})\nonumber\\
&&-m_{i+1/2}\frac{\partial\Phi(r_{i+1/2})}{\partial r_{i+1/2}},
\end{eqnarray}
\begin{equation}
\frac{dJ_{i+1/2}}{dt}=0,
\end{equation}
\begin{equation}
J_{i+1/2}\equiv\frac{\partial L}{\partial \dot{\theta}_{i+1/2}},
\end{equation}
where $J_{i+1/2}$ is the angular momentum defined at a cell-boundary $r_{i+1/2}$.

\section{Correction terms in self-gravity}\label{sec:ap2}
In this section, we introduce the correction terms in self-gravity. In the following, we consider the self-gravity exerted on $r_{i+1/2}$ as an example.

We introduce gravity from $r_{i}$ and $r_{i+1}$ as the correction terms, which represent gravity from fluid around $r_{i+1/2}$. As an example, we describe the way to calculate the gravity from the fluid at $r=r_{i}$. We divide the mass $m_{i}$ as follows. First, we define $m_{-,a}$ and $m_{-,b}$ as the masses in $r_{i}\le r \le r_{i+1/2}$ and $r_{i-1/2}\le r \le r_{i}$, respectively.
\begin{eqnarray}
&m_{-,a}&\equiv \pi\left[r_{i+1/2}^2-r_{i}^2 \right]\Sigma_{i},\\
&m_{-,b}&\equiv \pi\left[r_i^2-r_{i-1/2}^2 \right]\Sigma_{i}.
\end{eqnarray}
With these masses, we calculate masses defined at $r=r_{i}$, $r_{i-1/2}$ as follows (see figure \ref{fig:ap2f1}):
\begin{figure}[ht!]
%\centering
\begin{center}
%\begin{center}
%\includegraphics[bb=0 0 350 200,clip,width=\hsize]{./f18.pdf}
%\includegraphics[bb=0 0 350 200,clip,width=8cm]{./f18.pdf}
%\FigureFile(8cm,8cm){f2.eps}
\includegraphics[clip,width=8cm]{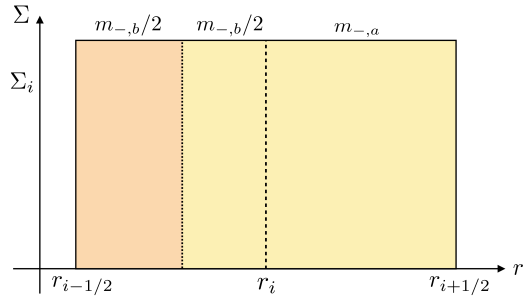}
\end{center}
\caption{Way of mass distribution used when we calculate the correction terms for the self-gravity. We only show the region $r\le r_{i+1/2}$. The mass defined at $r=r_{i+1/2}$ is given by $m_- = m_{-,a}+m_{-,b}/2$, where $m_{-,a}\equiv \pi\left[r_{i+1/2}^2-r_{i}^2 \right]\Sigma_{i}$ and $m_{-,b}\equiv \pi\left[r_i^2-r_{i-1/2}^2 \right]\Sigma_{i}$. We change $m_{i-1/2}$ into $\left[m_{i-1}+m_{-,b}\right]/2$. In the outer region $r>r_{i+1/2}$, we calculate the mass $m_+$ defined at $r=r_{i+1}$ in the same way, and we correct $m_{i+3/2}$ with equation (\ref{eqmi3-2}).\label{fig:ap2f1}}
\end{figure}
\begin{eqnarray}
&m_-& = m_{-,a}+\frac{m_{-,b}}{2},\\
&m_{i-1/2}&=\frac{m_{i-1}+m_{-,b}}{2}.
\end{eqnarray}
We also calculate the masses $m_{+}$ and $m_{i+3/2}$ defined at $r=r_{i+1}, r_{i+3/2}$ respectively using $m_{+,a}$ and $m_{+,b}$ as follows:
\begin{eqnarray}
&m_{+,a}&\equiv \pi\left[r_{i+1}^2-r_{i+1/2}^2 \right]\Sigma_{i+1},\\
&m_{+,b}&\equiv \pi\left[r_{i+3/2}^2-r_{i+1}^2 \right]\Sigma_{i+1},\\
&m_+&=m_{+,a}+\frac{m_{+,b}}{2},\\
&m_{i+3/2}&=\frac{m_{i+2}+m_{+,b}}{2}.\label{eqmi3-2}
\end{eqnarray}
With these masses, we calculate the gravity acting at $r=r_{i+1/2}$ from $r=r_{i-1/2}$, $r_{i}$, $r_{i+1}$, and $r_{i+3/2}$ in the same way as equation (\ref{sggrav}).
%\begin{figure}[ht!]
%\centering
%%\begin{center}
%%\includegraphics[bb=0 0 350 200,clip,width=\hsize]{./f18.pdf}
%\includegraphics[bb=0 0 350 200,clip,width=8cm]{./f18.pdf}
%\caption{Way of mass distribution used when we calculate the correction terms for the self-gravity. We only show the region $r\le r_{i+1/2}$. The mass defined at $r=r_{i+1/2}$ is given by $m_- = m_{-,a}+m_{-,b}/2$, where $m_{-,a}\equiv \pi\left[r_{i+1/2}^2-r_{i}^2 \right]\Sigma_{i}$ and $m_{-,b}\equiv \pi\left[r_i^2-r_{i-1/2}^2 \right]\Sigma_{i}$. We change $m_{i-1/2}$ into $\left[m_{i-1}+m_{-,b}\right]/2$. In the outer region $r>r_{i+1/2}$, we calculate the mass $m_+$ defined at $r=r_{i+1}$ in the same way, and we correct $m_{i+3/2}$ with equation (\ref{eqmi3-2}).\label{fig:ap2f1}}
%\end{figure}

\section{Interpolation function for the piecewise exact solution}\label{sec:ap3}

In this appendix, we explain an interpolation of physical quantity when we use the piecewise exact solution. Since we interpolate the quantity of both gas and dust with same way, we omit index $\dst$ and $\mathrm{g}$, and use radial velocity $u_{r,i+1/2}$ and specific angular momentum $l_{i+1/2}$. The interpolation function of the radial velocity $u_r(r)$ and the specific angular momentum $l(r)$ in $r_{i}\le r<r_{i+1}$ are the following equations (\ref{int_vr}) and (\ref{int_l}). The interpolation function of $l$ gives an exact value when the disk rotates with the Keplerian velocity:
\begin{equation}
u_r(r) = a(i)r+b(i),\label{int_vr}
\end{equation}
\begin{equation}
l(r) = \sqrt{r}\left(c(i)r+d(i)\right),\label{int_l}
\end{equation}
where the coefficients $a(i)$ and $c(i)$ are defined as 
\begin{eqnarray}
a(i)&=&\frac{u_{r,i+3/2}-u_{r,i-1/2}}{r_{i+3/2}-r_{i-1/2}}\;\;\;\; (i=2,3,...,N-1),\\
a(1)&=&a(2),\;\;a(N)=a(N-1),
\end{eqnarray}
\begin{eqnarray}
c(i)&=&\frac{l_{i+3/2}/\sqrt{r_{i+3/2}}-l_{i-1/2}/\sqrt{r_{i-1/2}}}{r_{i+3/2}-r_{i-1/2}}\\ 
&&\;\;\;\;\;\;\;\;\;\;\;\;\;\;\;\;\;\;\;\;\;\;\;\;\;\;\;\;\;\;\;\;\;\;\;\;\;\;\;\;\;\;\;\;\;\;(i=2,3,...,N-1),\nonumber\\
c(1)&=&c(2),\;\;c(N)=c(N-1).
\end{eqnarray}
We define $b(i)$ and $d(i)$ to satisfy the following relations:
\begin{eqnarray}
\int_{R(i)}^{r_{i+1/2}}2\pi r \Sigma_iu_r(r)dr&+&\int^{R(i+1)}_{r_{i+1/2}}2\pi r \Sigma_{i+1}u_r(r)dr\nonumber\\
&&=m_{i+1/2}u_{r,i+1/2}
\end{eqnarray}
\begin{eqnarray}
\int_{R(i)}^{r_{i+1/2}}2\pi r \Sigma_i l(r)dr&+&\int^{R(i+1)}_{r_{i+1/2}}2\pi r \Sigma_{i+1}l(r)dr\nonumber\\
&&=J_{i+1/2}
\end{eqnarray}
where $R(i)$ and $R(i+1)$ correspond to positions where the masses in the $i$-th and the $(i+1)$-th cells are equally divided, which is defined as
\begin{eqnarray}
R(1)&=&\sqrt{r_{3/2}^2-\frac{m_{1}}{2\pi\Sigma_1}}\\
R(i)&=&\sqrt{\frac{m_{i}}{2\pi\Sigma_i}+r_{i-1/2}^2}\;\;\;\; (i=2,3,...,N),
\end{eqnarray}

%%%
% See the manual for the detail.
%%%

\end{document}